# *In situ* stimulation of self-assembly tunes the elastic properties of interpenetrated biosurfactant-biopolymer hydrogels

Chloé Seyrig,[a] Alexandre Poirier,[a] Thomas Bizien,[b] Niki Baccile[a]

[a] Sorbonne Université, Centre National de la Recherche Scientifique, Laboratoire de Chimie de la Matière Condensée de Paris , LCMCP, F-75005 Paris, France

[b] Synchrotron SOLEIL, L'Orme des Merisiers Saint-Aubin, BP 48 91192 Gif-sur-Yvette Cedex

* Corresponding author:
Dr. Niki Baccile
E-mail address: niki.baccile@sorbonne-universite.fr
Phone: +33 1 44 27 56 77

**Abstract**

Hydrogels are widespread soft materials, which can serve a wide range of applications. The control over the viscoelastic properties of the gel is of paramount importance. Ongoing environmental issues have raised the consumer's concern towards the use of more sustainable materials, including hydrogels. However, are greener materials compatible with high functionality? In a safe-by-design approach, this work demonstrates that functional hydrogels with *in situ* responsivity of their elastic properties by external stimuli can be developed from entirely "sustainable" components, a biobased amphiphile and biopolymers (gelatin, chitosan and alginate). The bioamphiphile is a stimuli-responsive glycolipid obtained by microbial fermentation, which can self-assemble into fibers, but also micelles or vesicles, in water under high dilution and by a rapid variation of the stimuli. The elastic properties of the bioamphiphile/biopolymer interpenetrated hydrogels can be modulated by selectively triggering the phase transition of the glycolipid and/or the biopolymer inside the gel by mean of temperature or pH.

## Introduction

Classical hydrogels are intended as physically or chemically cross-linked single polymer networks with a broad set of applications, from biomedical to food science.[1–5] They generally offer very good mechanical properties and, when the polymer is obtained by natural sources, their biocompatibility allows use in human-related applications. However, their functions could be limited due to a lack of dynamic and structural complexity within the hydrogels. Advanced engineering of parameters such as mechanics and spatially/temporally controlled release of (bio)active molecules, as well as manipulation of multiscale shape, structure, and architecture, could significantly widen their applications.[1,2]

One strategy increasingly employed to address such challenges is the incorporation of a second polymer, resulting in interpenetrated network (IPN) hydrogels, accumulating the properties of each component.[1] Such strategy is targeted by many reviews[6–8] with a large variety of compounds, stimuli (glucose, pH, magnetism, enzyme, light, mechanics …), methods of preparation, experimental conditions. Van Esch *et al*., for example, evidenced that the catalyst-stimulated *in situ* gelation of a hydrogelator occurs faster (min *vs*. h) and results in materials displaying controllable stiffness in the 5–50 kPa range, depending on the concentration of the catalyst.[9]

Among a wide range of stimuli, pH-responsiveness has attracted increasing interest due to practical purposes. pH is a handy tool to release locally a bioactive component based on the acid-base conditions of various tissues targeted (e.g., tumor microenvironment, gastric fluid, colon).[10] Within this framework, however, the use of polymers is sometimes limited by their slow kinetics against stimuli responsivity. For this reason, recent research started to look at the benefits of low-molecular weight (LMW) amphiphiles, of which the self-assembly is reversible and the sol-gel transition fast. Recent work was then dedicated to study the interpenetration and responsivity of low molecular weight gelators (LMWG), either in combination with polymers gelators (PG), or self-combined.[11–19]

Adams *et al*. synthetized a multicomponent supramolecular gel, of which the unusual phase behavior, driven by two opposite ionizable pendant groups on different components, allows access to two different gel states both at acidic and basic pH.[11] A similar difference in pH sensitivity LMW systems was also studied by Piras and Smith,[13] who inversely switched "off" and "on" two networks, one containing an amine and the other an acid. Cornwell and Smith proposed a two-component hybrid system in which the contribution of each network and the self-sorting notion are put in evidence.[14] Despite these and other examples, the development of multifunctional hydrogels able to respond to multiple stimuli remains challenging. To achieve this, IPN hydrogels based on multiple networks could be designed to respond to different stimuli. Some temperature/pH dual responsive systems already exist and exhibit an efficient additional control over the delivery of therapeutic drugs and proteins.[18,19]

If the improvement of the mechanical properties in IPN systems under equilibrium conditions has been discussed before,[20–22] including our own work,[23] there has been no effort devoted to the dynamic control of the rheological properties through stimulation of phase, and state, transitions of the LMW network. To the best of our knowledge, phase transitions are rarely triggered inside the gel phase and poor attention was paid to their effect on the mechanical properties.[18,24]

In this regard, this work aims at developing a stimulable, synergistic, IPN composed of a LMWG and a PG, where the self-assembly of the LMWG, induced by a dynamic variation of a given external stimulus (pH or temperature), occurs in the hybrid gel matrix. The stimulated, reversible, sol-gel transition of the LMWG controls the elastic properties of the PG, promoting a prompt recovery of its elastic modulus. On the other hand, the PG improves the elastic properties of the LMWG, notoriously weaker and more sensitive to external perturbations than the PG. The second objective of this work is to show that such a complex soft system can entirely be developed from functional bio-based molecules, a biopolymer and a bioamphiphile, without the need of any otherwise classical chemical engineering approach.

The selected bioamphiphile, G-C18:1 (Figure 1a), is a fermented bolaform glycolipid composed of a single *β*-D-glucose hydrophilic headgroup and a C18:1 fatty acid tail. Due to its double amphiphilic nature and its free-standing COOH group, G-C18:1 is negatively-charged under alkaline conditions, displaying a complex multiphase behaviour under diluted conditions in water.[25] When negatively-charged, G-C18:1 forms a micellar phase[26,27] in the presence of $Na^+$ and a fiber phase with $Ca^{2+}$ (Figure 1b).[28,29] Reducing pH from alkaline to acidic drives a micelle-to-wormlike-to-vesicle-to-lamellar phase transition (Figure 1b).[26,27] Recent work has even shown its ability to enrich its own phase diagram by strongly binding to cationic polyelectrolytes (e.g., poly-L-lysine or chitosan), but not to amphoteric polymers, like gelatin, under like-charge conditions.[23,30,31] Considering the importance of the electrostatic interactions between surfactants and polymers,[32] the selected biopolymers for this work are gelatin,[33] chitosan and alginate (Figure 1a).[34,35] Not only they have an amphoteric, positive and negative character, respectively, they are also among the most studied and commercially-exploited biopolymers, thus opening formulation perspectives of this work in the fields of cosmetics or medical science.

The present work shows that a safe-by-design approach can be successfully employed to prepare soft IPN composed of a PG and a LMWG with mutual synergy. Combination of rheology and small angle X-ray scattering using synchrotron radiation allows pH- and temperature-responsive *in situ* experiments, which show that bioamphiphiles and biopolymers can interpenetrate into an orthogonal network, where the sol-gel transition of each component can independently be stimulated. Besides modulating the overall hydrogel's elastic properties, the effect of orthogonality is such that as a function of given conditions of pH or temperature, either the bioamphiphile or the biopolymer prevent the overall hybrid's gel collapse, which would occur in single components hydrogels.

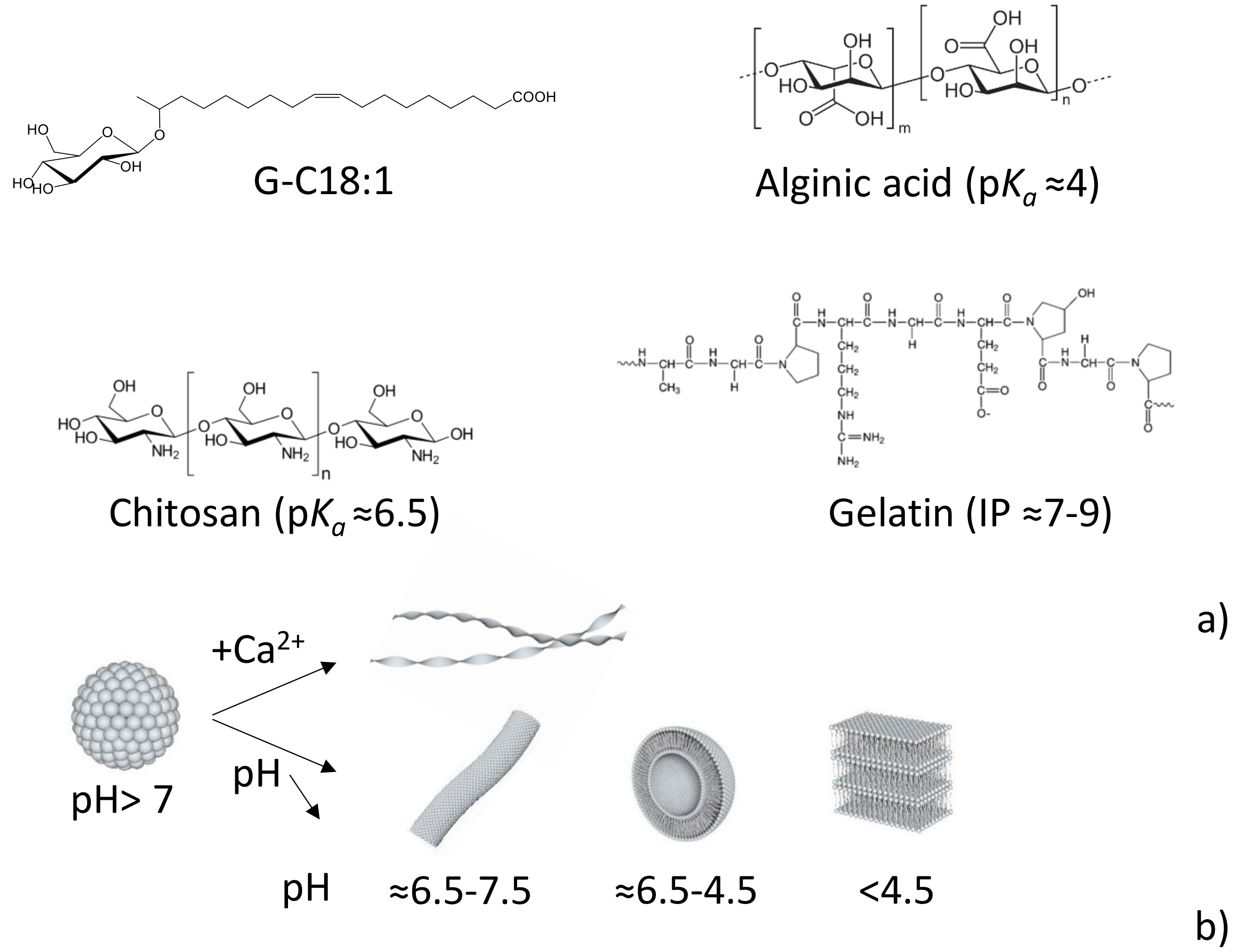

**Figure 1 – a) Chemical formulas of G-C18:1, alginic acid, chitosan and gelatin. IP: isoelectric point. b) Known phase behavior in water at room temperature (C< ~5 wt%) of G-C18:1. Micelles form at basic pH (counterions: $Na^+$) and undergo a wormlike-to-vesicle-to-lamellae transition.[26,27] Crystalline fibers form from the micellar phase at pH> 7 by adding $Ca^{2+}$.[28,29,36]**

**Experimental section**

*Chemicals*

Glycolipid biosurfactant, glucolipid G-C18:1, is purchased from the Bio Base Europe Pilot Plant, Gent, Belgium, lot N° APS F06/F07, Inv96/98/99 and used as such. The monounsaturated glucolipid G-C18:1 ($M_w$ = 460 g.mol$^{-1}$) contains a β-D-glucose unit covalently linked to oleic acid. The molecule is obtained by fermentation from the yeast *Starmerella bombicola ΔugtB1* according the protocol given before.[37] The molecular purity of G-C18:1 exceeds 95%. According to the specification sheet provided by the producer, the batch (99.4% dry matter) is composed of 99.5% of G-C18:1, according to HPLC-ELSD chromatography data. NMR analysis was performed elsewhere.[27] The three polymers used in this work, gelatin (type A, from porcine skin, $M_w$ ≈50–100 kDa, isoelectric point 7–9), alginic acid (from brown algae, medium viscosity, $M_w$ ≈20–240 kDa, p$K_a$ ≈4) and chitosan (high molecular weight, HMW, from shrimp shell, practical grade, $M_w$ ≈190–375 kDa, p$K_a$ ≈6.5), are purchased from Sigma Aldrich. Glucono-δ-lactone, GDL, is purchased at Aldrich.

*Preparation of the hydrogels*

*Stock solutions*. The G-C18:1 stock solution at concentration, $C_{G\text{-}C18:1}$= 40 mg/mL, is prepared by dissolving the G-C18:1 powder in the appropriate volume of milli-Q grade water at alkaline pH (generally 8 or 10), adjusted with 5-10 µL of NaOH 5 or 1 M solution. The stock solutions for each biopolymer are prepared as follows.

*Gelatin*: 80 mg of gelatin powder is dispersed in 2 mL of milli-Q water, for a concentration of $C_{gelatin}$= 40 mg/mL. The gelatin stock solution is vortexed and set in the oven at 50°C. Once the solution is homogeneous, pH is increased (generally 8 or 10) with a few µL of a 0.5 M - 1 M NaOH solution. The protocol to prepare the alginate and chitosan stock solutions is given below. One should note however that, due to improved handling of the chitosan and alginate gels, lower concentrations of the stock solutions (20 mg/mL, half the one of gelatin) were employed. It was found that lower concentration of the stock solution only affects the magnitude of the initial gel strength, but not the phenomena described in this work.

*Chitosan*: 200 mg of chitosan dispersed in 10 mL of 0.1 M acetic acid aqueous solution for a concentration C= 20 mg/mL. For an optimal solubilisation, the chitosan stock solution is stirred during one day before use. The pH of the acidic chitosan stock solution is increased above about 8. Typically, for 1 mL, one adds 5-10 µL of a 5 M or 1 M NaOH solution. The solution, initially viscous at acidic pH, is vigorously stirred and vortexed upon pH increase to obtain a heterogeneous gel.

*Alginate*: 200 mg of alginic acid powder is dispersed in 10 mL of Milli-q water for a concentration C= 20 mg/mL and stirred until complete solubilization. The magnetic stirrer usually sticks upon water addition; in this case, a manual help may be required to improve stirring. For a typical volume of 10 mL, pH is then increased to 8 or 10 with a 1-5 µL of a 5 M or 1 M NaOH solution under stirring. Stirring and vortexing are eventually necessary to obtain a homogeneous alginate solution. Additional protocol details are given below for each biopolymer.

*Fiber G-C18:1 hydrogels, {F}G-C18:1.* Preparation of fibrillar gels of G-C18:1 were described elsewhere.[28,29,36] Typically, for a 1 mL sample, one adds a given volume of a $CaCl_2$ solution (1 M, $V_{CaCl2}$= 33.5 µL, $[Ca^{2+}]$ = 33.5 mM) to 1 mL of the G-C18:1 stock solution for a total $[Ca^{2+}]$ : [G-C18:1]= 0.7 molar ratio. The final solution is stirred and a gel is obtained after resting a few hours at room temperature.

*{F}G-C18:1/gelatin gels.* A volume of 500 µL of the gelatin stock solution (pH, 8 or 10, is given in the figures' legend) is mixed with 500 µL of either water (reference) or G-C18:1 stock solution (sample). For a typical volume of 1 mL, $CaCl_2$ solution (1 M, $V_{CaCl2}$= 33.5 µL, $[Ca^{2+}]$ = 33.5 mM) is manually added for a total $[Ca^{2+}]$ : [G-C18:1]= 0.7 molar ratio. The final solution is stirred and a gel is obtained after resting a few hours at room temperature. For the pH-dependent using GDL, rheo-SAXS experiment, these values are divided by a factor two. The exact values are given in the main text below each experiment.

*{F}G-C18:1/chitosan gels*. Due to the heterogeneity of the chitosan gel at basic pH (8 or 10, given in the figures' legend), 500 mg of the chitosan gel are weighted and mixed with 500 µL of a glycolipid solution under vigorous stirring and vortexing. To this mixture, 33.5 mM ($V_{CaCl2}$= 67 µL) of a $CaCl_2$ solution ($[Ca^{2+}]$ = 1 M) are added, followed by further mixing. The final concentration of $CaCl_2$ in the sample is 33.5 mM, for a total $[Ca^{2+}]$: [G-C18:1] = 0.7 molar ratio. A homogeneous gel is obtained after resting a few hours at room temperature. For the pH-dependent using GDL, rheo-SAXS experiment, these values are divided by a factor two. The exact values are given in the main text below each experiment.

*{F}G-C18:1/alginate gels.* 500 µL of the alginate viscous stock solution are added either to 500 µL water (reference) or to 500 µL of the G-C18:1 stock solution under stirring. Final pH, 8 or 10, is given in the figures' legend. A volume of $V_{CaCl2}$= 50 µL of a $CaCl_2$ solution ($[Ca^{2+}]$= 1

M) is added, for a final $CaCl_2$ concentration of 50 mM and $[Ca^{2+}]$ : [G-C18:1]= 0.9 molar ratio and $[Ca^{2+}]$ : [alginate]≈ 0.0015 molar ratio. The final solution is magnetically stirred for several hours to obtain a homogeneous gel. For the pH-dependent using GDL, rheo-SAXS experiment, these values are divided by a factor two. The exact values are given in the main text below each experiment.

Table 1 summarizes the reference, sample and stock solution concentrations in wt%. These values are divided by a factor two for the pH-dependent experiments using GDL.

**Table 1 – Concentration of stock solutions, volumes from stock solutions and final concentration of the samples. Please refer to the *stock solution* paragraph for the justification in employing reduced chitosan and alginate concentrations.**

| | G-C18:1 | gelatin | alginate | chitosan |
|---|---|---|---|---|
| $C_{stock\ solution}$ / wt% | 4 | 4 | 2 | 2 |
| V / mL | 0.5 | 0.5 | 0.5 | 0.5 |
| $C_{sample}$ / wt% | 2 | 2 | 1 | 1 |

*Rheology*

Viscoelastic measurements are carried out using an Anton Paar MCR 302 rheometer equipped with parallel titanium or stainless steel sandblasted plates (diameter = 25 mm, gap = 1 mm). Unless otherwise stated, all experiments are conducted at 25 °C, whereas the temperature is controlled by the stainless steel lower Peltier plate. During the experiments, the measuring geometry is covered with a humidity chamber to minimize water evaporation. To investigate the pH-dependence of the mechanical properties, samples containing {F}G-C18:1 and the biopolymer are mixed with the appropriate amount (values are given in the main text) of glucono-δ-lactone, GDL, and immediately vortexed during 20 s. Volumes have been doubled and concentrations have been adapted (divided by two) for an easier dispersion of GDL powder. Half of the sample is immediately loaded on the bottom plate, while the pH is monitored automatically on the other half. Dynamic oscillatory and time sweep experiments are performed by applying a constant oscillation frequency (f = 1 Hz) and a shear strain ($\gamma$= 0.1 %) within the linear viscoelastic regime (LVER).

*pH monitoring*

*In situ* pH monitoring after addition of GDL is performed using a Mettler Toledo microelectrode connected to a Hanna scientific pH-meter, model HI 5221. The pH meter is connected to a computer, equipped with the fabricant's software [HI 92000, version 5.0.28]. The frequency of pH recording is 10 $s^{-1}$.

*Rheo-SAXS*

Experiments coupling rheology and SAXS are performed at the SWING beamline of the Soleil synchrotron facility (Saint-Aubin, France) during the run N° 20200532, using a beam energy of 12.00 keV and a sample-to-detector distance of 1.65 m. Tetradecanol ($d_{(001)}$ = 39.77 Å) is used as the q-calibration standard. The signal of the EIGERX 4M 2D detector (75 µm pixel size) is integrated azimuthally with Foxtrot software to obtain the I(q) spectrum ($q = 4\pi \sin \theta/\lambda$, where 2θ is the scattering angle) after masking systematically defective pixels and the beam stop shadow. Data are not scaled to absolute intensity. A MCR 501 rheometer (Anton Paar, Graz, Austria) equipped with a Couette polycarbonate cell (gap= 0.5 mm, volume, ≈ 1.35 mL) is coupled to the beamline and controlled through an external computer in the experimental hutch using the Rheoplus/32 software, version 3.62. The experiments are performed in a radial configuration, where the X-ray beam is aligned along the center of the Couette cell. The rheology and SAXS acquisitions are triggered manually with an estimated delay of less than 2 s. Due to standard compulsory security procedures required at the beamline, the first rheo-SAXS experimental point is systematically acquired with a delay of about 2–3 minutes with respect to the rheometer.

The experimental procedure for the rheo-SAXS is the same as the one described in the rheology section: a solution of {F}G-C18:1 and the biopolymer are mixed in a vial. To the vial, the appropriate amount (values are given in the main text) of glucono-δ-lactone, GDL, is added and the vial is vortexed during 20 s. For an easier dispersion of GDL powder, concentrations of G-C18:1 and biopolymer are divided by a factor two, with respect to the values in Table 1. Immediately after GDL addition, half of the sample solution is loaded in the couette cell, while the other half is used for pH monitoring using the automatic pH acquisition. Rheology is performed under oscillatory conditions within the linear viscoelastic regime (LVER, f = 1 Hz, $\gamma$= 0.1 %) as a function of time.

*Scanning electron microscopy (SEM)*

SEM was done on a Hitachi S3400N instrument working with an accelerated voltage of 10 kV. Samples were observed as such, without metallization. Images were magnified 500, 2000 and 5000 times. Acquisition time per image was 40 s.

*Solid-state magic angle spinning (MAS) nuclear magnetic resonance (ss-NMR).*

$^{1}H$ and $^{13}C$ ss-NMR spectroscopy was performed on a Bruker Avance III 700 MHz (16.4 T) spectrometer. Spectra were recorded with a 3.2 mm zirconia rotor as sample holder spinning at MAS rate, $\nu_{MAS}$ = 20 kHz. The chemical shift reference for both $^{1}H$ and $^{13}C$ was adamantane ($\delta_{1H}$ = 1.86 ppm; $\delta_{13C}$= 38.52 and 24.47 ppm). Both $^{1}H$ and $^{13}C$ spectra were recorded using direct polarization ($^{1}H$: *zg* pulse scheme; $^{13}C$: *hpdec* pulse scheme in Bruker's library), with a Spinal64 decoupling scheme during the $^{13}C$ signal acquisition. Recycle delay was 3 s for $^{1}H$ acquisition and 5 s for $^{13}C$ acquisition, while number of transient was 4 ($^{1}H$) and 960 ($^{13}C$). $^{1}H$ and $^{13}C$ nutation (90°) frequencies were 6.3 μs for both nuclei. An exponential line broadening of 100 Hz was used during $^{13}C$ signal processing. The TopSpin software, 3.6.2 version, was used for data acquisition and treatment (Fourier transform, phasing, line broadening, baseline correction).

*Sample preparation for SEM and ss-NMR experiments.*

All samples were prepared using the protocol described above and with concentrations given in Table 1. More specifically, {F}G-C18:1 and biopolymer (gelatin, 2 wt%; alginate, 1 wt%, [$Ca^{2+}$]= 25 mM) control samples were prepared at pH 10 and pH 4, plunged into liquid nitrogen and freeze-dried during 48 h.

For the IPN samples, a 5 mL solution of {F}G-C18:1/gelatin at pH 10 was split into four fractions of 1.5 mL each, while a 10 mL solution of {F}G-C18:1/alginate at pH 10 was split into four fractions of 2.5 mL each. Each fraction at pH 10 containing both biopolymers were plunged into liquid nitrogen and freeze-dried during 48 h, while GDL ($C_{GDL}$: 10 mg/mL for {F}G-C18:1/gelatin and 20 mg/mL for {F}G-C18:1/alginate) was added to the remaining 6 fractions. pH was monitored for each fraction during acidification and as soon as pH hit the desired values (6.7, 6 and 4) the corresponding sample was also plunged into liquid nitrogen and freeze-dried during 48 h. The choice of the pH was done according to the data collected during the pH-resolved *in situ* rheo-SAXS experiments, as discussed in the main text. In total, the following freeze-dried samples were collected and analyzed by SEM and ss-NMR: {F}G-C18:1 at pH 4 and 10, gelatin at pH 4 and 10, alginate at pH 4 and 10, {F}G-C18:1/alginate at pH 10, 6.7, 6 and 4, {F}G-C18:1/gelatin at pH 10, 6.7, 6 and 4.

**RESULTS**

G-C18:1 is known to have a multiphase behaviour according to pH or type of ion, as shown in Figure 1b.[25] In this work, {F}G-C18:1 refers to a stable, equilibrated, fiber hydrogel driven by adding $Ca^{2+}$ to the micellar phase of G-C18:1. The fibers, reported and characterized for the first time elsewhere,[28,29] are infinitely long crystalline threads having a cross section of about 10 nm.[29] The fibers promptly form a hydrogel in water at concentrations below 5 wt% (frequency and strain-dependent small amplitude oscillatory shear experiments are reported in Ref. [36]) and undergo a gel-to-sol (fiber-to-micelle) transition between 60°C and 63°C.[36] The effect of adding calcium on a G-C18:1 solution at basic pH is qualitatively shown in Figure 2: the G-C18:1 solution (2 wt%) at pH 10 is liquid, while the corresponding fiber phase, {F}G-C18:1, triggered by calcium itself,[29] forms a stable gel.

On the other hand, the biopolymers in this work were selected for their biobased nature, their well-known gelling properties at room temperature and at pH above 6 (Figure S 1) and their differences in terms of charge (chitosan: neutral/positive; alginate: negative neutral; protein: positive/negative). The micellar, vesicular or fibrillar forms of G-C18:1 was specifically studied under equilibrium conditions in combination with the biopolymers using both SAXS and small amplitude oscillatory rheology in a previous study.[23] It was found that the nature of the G-C18:1 phase has an impact on the elastic properties of the biopolymer's hydrogel: the fibrillar phase ({F}G-C18:1) systematically improves the hydrogel's strength, while micellar and vesicular phases soften the biopolymer's gel. In the specific case of {F}G-C18:1/alginate, $Ca^{2+}$ triggers gelation for both components individually and excess of calcium improves the elasticity of the IPN over each component taken individually. This suggested possible intramolecular cross-linking between the carboxylate groups of G-C18:1 and alginate.[23] The effect of mixing the biopolymer and {F}G-C18:1 at pH 10 is qualitatively shown in Figure 2. The interpenetrated gels become more opaque and stiffer, as they are resistant to shearing, by hand or vortexing. The effect of adding a source of calcium to the G-C18:1/alginate (turning into {F}G-C18:1/alginate (pH 10)) system is particularly self-evident by the sol-to-gel transition shown by the corresponding image in Figure 2.

Given the above, the elastic properties of the specific self-assembled fiber-biopolymer network are now explored under non-equilibrium conditions, after triggering phase transitions of G-C18:1, using an *in situ* approach. The structure and mechanical properties of {F}G-C18:1/biopolymer IPN hydrogels, initially set at equilibrium at room temperature and alkaline pH (pH 8 or 10, noted in the figures' legends),[23] are monitored *in situ* using a rheo-SAXS apparatus. The properties are modified by a change in temperature or pH under non-equilibrium conditions. Temperature is modified inside the couette cell of the rheometer while pH is modified at room temperature by mean of glucono-δ-lactone (GDL), a small hydrophilic sugar which spontaneously hydrolyzes in water into gluconic acid.

GDL has been reported several times in association to LMWG, alginate and even formulation of LMWG and alginate. When employed in the presence of LMWG, GDL is known not to interfere neither with fibrillation nor with gelation, if compared to more classical HCl. This aspect was specifically verified by us and others[38–43] and confirmed by SAXS in this work, as discussed later in the manuscript. Similar conclusions were established when GDL is used with alginate and formulations of alginate and LMWG. Despite being a weak acid, GDL is actually employed as a trigger, and not a competitor, of alginate gelation[44,45] in the presence of $CaCO_3$, even in complex media containing a LMWG.[44]

*Use of pH to control the IPN properties*

pH controls the micelle-to-vesicle-to-lamellar phase behavior in G-C18:1 solutions and the sol-gel transition in {F}G-C18:1. These were shown extensively in Ref. 26,27 and 36, respectively. Qualitatively, the gel-to-sol transition triggered by GDL on {F}G-C18:1 (pH 10) gel is shown in the images in Figure 2. On the other hand, pH has no effect on the gel properties of the biopolymers, except for the magnitude, as shown on the controls measured at pH 8 and 6. According to experiments presented Figure S 1, the absolute magnitude of storage (G') and loss (G'') moduli depend on the biopolymer : 20-50, 50-100 and ≈2000 Pa for gelatin, alginate and chitosan respectively, but, as a general trend, pH does not sensibly affect G' and G'', except for chitosan, known to be more soluble at acidic pH.

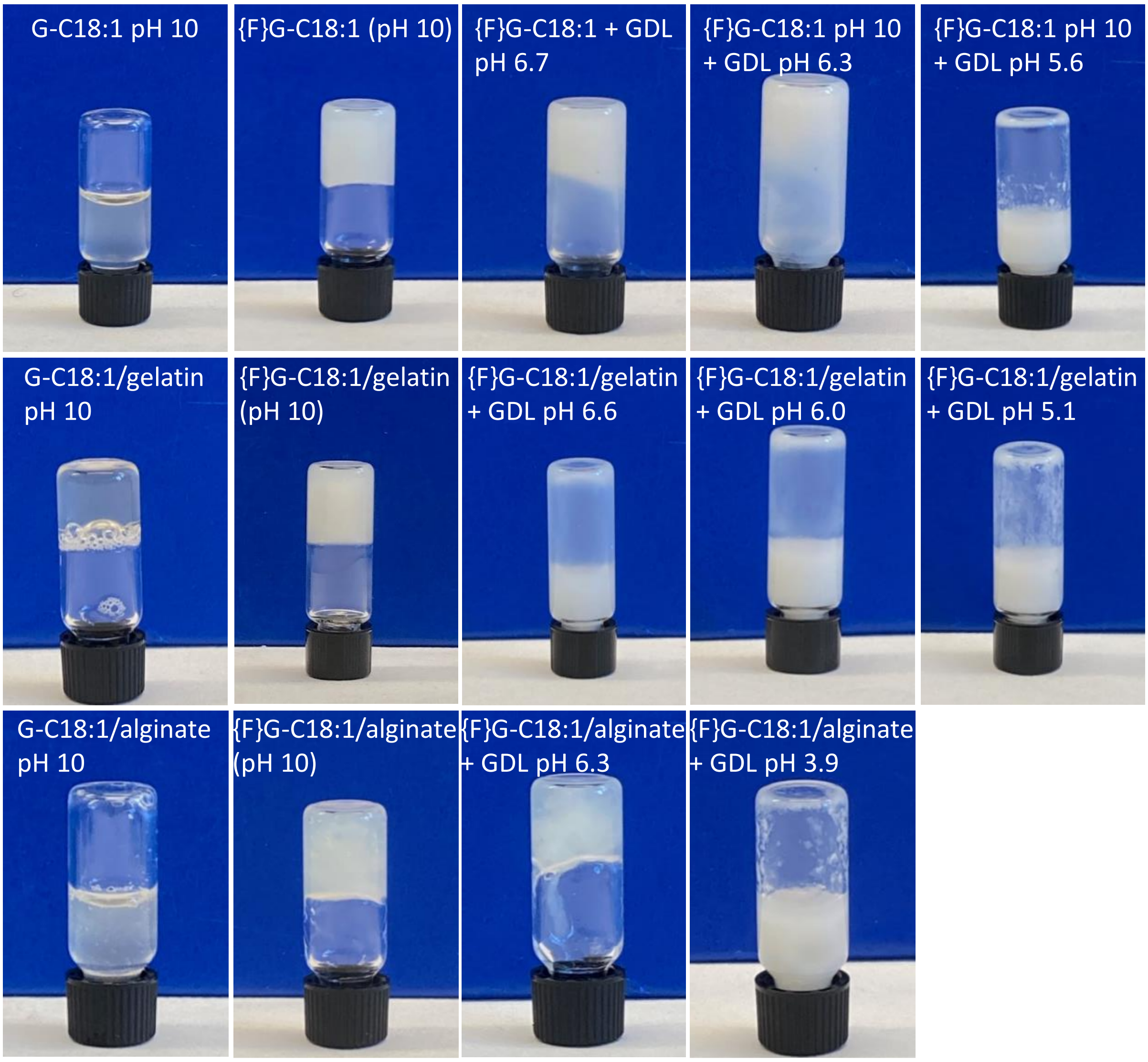

**Figure 2 – Images of G-C18:1, G-C18:1/gelatin and G-C18:1/alginate aqueous solutions (composition given in Table 1) at pH 10 before and after adding a source of $Ca^{2+}$, and after adding GDL. When calcium is added, G-C18:1 (micellar phase) becomes {F}G-C18:1 (fiber phase).**[29]

Time-dependent rheological measurements of all hybrid {F}G-C18:1/biopolymer hydrogels at pH 10 before pH variation are acquired on the SAXS-coupled rheometer before triggering the pH variation, while the corresponding frequency-dependent experiments are reported in Ref. [23]. The *ex situ* panel in Figure 3a, Figure 4a and Figure 5a shows that hybrid gels made of {F}G-C18:1 and gelatin, alginate and chitosan, respectively, have a storage modulus around 100 Pa at pH 10. After adding GDL to the hybrid hydrogels at pH 10, the sample is split in two halves for two parallel *in situ* measurements: the mechanical and structural properties are collected by rheo-SAXS on one half, while pH is measured using an automated pH meter on the other half. The results are combined in Figure 3, Figure 4 and Figure 5 for gelatin, alginate and chitosan, respectively. Due to the unavoidable security constraints of the SAXS beamline, a lag time of about 2 min systematically precedes rheo-SAXS acquisitions.

The elastic properties of the hybrid {F}G-C18:1/gelatin hydrogel are lost when adding GDL (Figure 2), with G′ dropping to ~5 Pa during acidification at pH 7, below which the gel becomes a liquid, characterized by a loss modulus higher than the storage modulus (Figure 3a). Figure 3b shows the concomitant SAXS signature (profiles *(1)* to *(3)* correlate SAXS and rheology in rheo-SAXS experiment), identifying the structural features of the gels. SAXS highlights a decrease in intensity of the peaks at q= 0.25 $Å^{-1}$ and q= 0.31 $Å^{-1}$, both characteristics of the crystalline structure of {F}G-C18:1 fibers, discussed elsewhere,[29] visible at pH 10. The peaks progressively disappear at more acidic pH, suggesting a structural perturbation of the system around pH ≈7. The SAXS profile at pH 6.47 (Figure 3b) should be compared to the SAXS signature of G-C18:1 under comparable conditions of dilution and pH. It is important to note that the SAXS patterns of the GDL-containing hybrid {F}G-C18:1/gelatin hydrogel, and recorded between pH 10 and pH 7, are comparable with the SAXS patterns of the GDL-free {F}G-C18:1 and {F}G-C18:1/gelatin hydrogel, previously shown in Ref. [23]. This fact confirms that GDL does not have any influence on the structure of the gels, as expected from literature.[38–43]

The pH-dependent self-assembly of G-C18:1 has been reported before[26] and the corresponding SAXS profiles, each associated to a given structure, are reproduced in Figure S 2a. Superposing the signal of {F}G-C18:1/gelatin at pH 6.47 shows the similarity between the latter and the signal of G-C18:1 alone at pH 6.6, representative of a transitory wormlike micelle phase.[26] Although model-dependent fitting is often employed to extract quantitative parameters from SAXS profiles, this approach is not adapted to the present study, due to the superposition of more than one SAXS signal, thus making the fitting process not only complex and time-consuming, but also unreliable. The SAXS signal of G-C18:1 was fitted with two concomitant models, core-shell ellipsoid and core-shell bicelle form factors, to account for the micellar and vesicle phase, respectively.[26] On the other hand, the fiber phase of G-C18:1 is even more complex, as it requires two additive form factor models (core-shell parallelepiped model, lamellar paracrystal) and a structure factor.[29] The above require a fitting process with no less than twenty independent parameters, of which at least half are critical. To this system, one should add the contribution of the biopolymer in the present work, most likely using a classical flexible cylinder model,[46] which requires seven additional independent parameters. Overall, this process eventually requires not less than twenty-five independent parameters, thus making the result highly uncertain and questionable.

However, a simple, classical, model-independent approach consisting in measuring the slope (log-log scale) in the low-q region confirms the wormlike micellar structure for {F}G-C18:1/gelatin at pH 6.47. Cylindrical and flat morphologies scatter with a -1 and -2 dependency of the wavevector at low-q,[47] while wormlike micelles are generally associated to intermediate exponents, generally in the order of -1.6[48,49] for their similarity with swollen polymer chains.[50,51] This is schematically illustrated in Figure S 2b, where three SAXS profiles are simulated (SasView 3.1.2 software) using a sphere, cylinder and flexible cylinder model,[46] with a radius of 18-20 Å (detailed list of model parameters are given in the caption of Figure S 2b). Experimentally, slope values in the order of -1.6 are measured for G-C18:1 at pH 6.6 (Figure S 2a) and for which the existence of wormlike micelles was demonstrated by cryo-TEM in a previous work.[26] Similarly, Figure S 3 shows a similar -1.6 slope for {F}G-C18:1/gelatin in the 6.4-6.6 pH range, while Figure S 2a shows how the profile of {F}G-C18:1/gelatin is almost superimposable to the one of G-C18:1 alone at pH 6.6, strongly suggesting that micelles are also formed in {F}G-C18:1/gelatin.

A closer comparison between rheology and SAXS clearly associates the loss of the mechanical properties to a loss in structure of {F}G-C18:1 fibers. The pH-dependent 2D SAXS contour plot centered around the fibers' structural diffraction peaks at q= 0.25 $Å^{-1}$ and q= 0.31 $Å^{-1}$ is superposed to the G'(pH) profile in Figure 3a, both synchronized within the same rheo-SAXS experiment. The loss in structural properties, marked as *(2)* and also visible by the naked eye (image *(2)* in Figure 3a), occurs in the same pH range, during which the structural peak disappears.

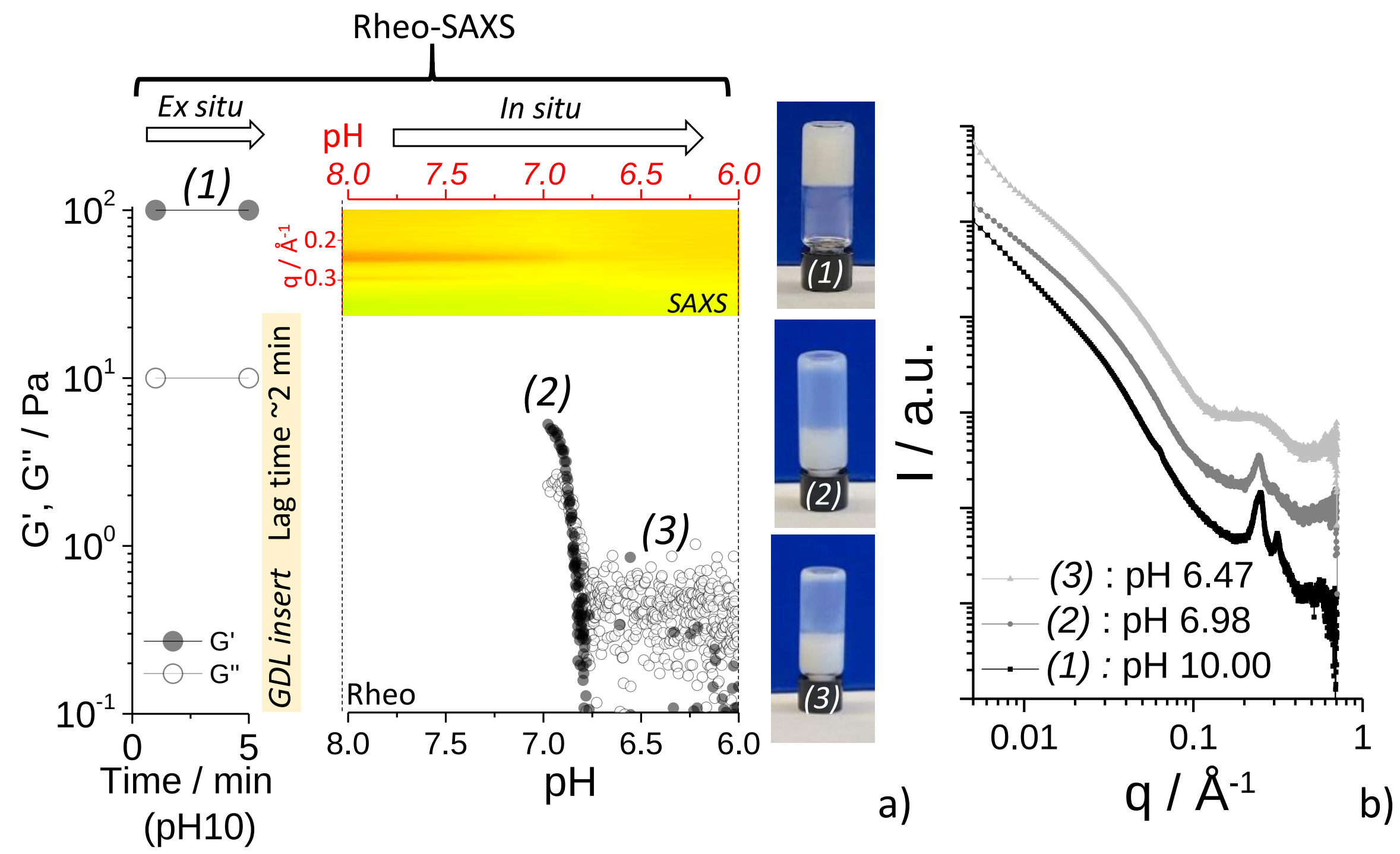

**Figure 3 – GDL-induced acidification of {F}G-C18:1/gelatin hydrogel. pH-resolved *in situ* rheo-SAXS experiment: a) mechanical properties (LVER, f= 1 Hz, γ= 0.1 %) and b) corresponding SAXS profiles. The 2D contour plot profile superposed to the G'(pH) evolution in a) corresponds to the full pH-dependent SAXS experiment, synchronized with rheology. $C_{GDL}$= 1 wt%, $C_{G\text{-}C18:1}$= 1 wt%, [$Ca^{2+}$] = 17.8 mM, $C_{gelatin}$= 1 wt%. The experimental details of the rheo-SAXS experiment are given in the experimental section. The control rheology experiment performed on gelatin at acidic and basic pH is given in Figure S 1a. The lag time in a) is due to the unavoidable security constraints of the SAXS beamline.**

After addition of GDL to the hybrid {F}G-C18:1/alginate hydrogels, the elastic modulus decreases with pH. Differently than gelatin, the {F}G-C18:1/alginate remains a gel in the explored pH range (the storage modulus is one order of magnitude above the loss modulus), to at least 5. This is qualitatively illustrated by the images of {F}G-C18:1/alginate + GDL in Figure 2, which show that the system is still gelled at pH 6.3. Compared to the alginate controls at pH 6 and 8 (Figure S 1b), the loss in elasticity at acidic pH is more pronounced by a factor three to four in the IPN system, illustrating the softening effect attributed to the gel-to-sol transition of G-C18:1, and contributing to an overall loss of the elastic properties at much more acidic pH (3.9, Figure 2).

The structural study performed by SAXS (Figure 4b) shows that the peaks (q= 0.25 and 0.31 $Å^{-1}$, profile (*1*) in Figure 4b) at pH 10, typical of the {F}G-C18:1 fiber phase, progressively disappear in favor of a wormlike phase in the pH region between 6 and 7. This is suggested by the SAXS profile *(2)* in Figure 4b. Its low-q slope is -1.56 (Figure S 3) and, similarly to the gelatin system discussed above, it is almost superimposable to G-C18:1 alone at pH 6.6 (Figure S 2a). Further decrease in pH provides SAXS curves (*3*) to (*5*), which display two diffraction peaks in a 1:2 relationship, typical of a lamellar structure. Effect of pH on {F}G-C18:1/alginate hydrogels identifies a wormlike-to-lamellar transition, the latter reported before and expected in the phase diagram of G-C18:1 alone (Figure 1, Figure S 2).[30,36] Interestingly, the vesicle phase, generally observed between the wormlike and lamellar phases for G-C18:1 alone (Figure 1, Figure S 2) below pH 6.5,[30,36] is not detected in the {F}G-C18:1/biopolymer systems, as also reported for calcium-containing G-C18:1 solutions prepared below pH 7.[36] The above is also well-illustrated by the pH-dependent 2D contour plot associated to the full rheo-SAXS experiment (2D SAXS panel superposed to its synchronized G'(pH) profile), which nicely shows that the fiber (pH> ~7) to lamellar (pH< ~6.5) phase transition occurs at the same time as the partial loss in elastic properties of the hybrid {F}G-C18:1/alginate hydrogels.

Finally, none of the effects described here are attributed to the presence of GDL. In fact, comparison of the SAXS profiles given in Figure 4b with the ones associated to GDL-free controls ({F}G-C18:1 and {F}G-C18:1/alginate), reported in Ref. [23], shows good agreement, thus excluding an impact of GDL neither on the structure of the fibers nor on the gelation properties.

All in all, the gel-to-sol transition of {F}G-C18:1 does not have as much effect on the elastic modulus of the {F}G-C18:1/alginate IPN when compared to gelatin. Nonetheless, this system nicely illustrates the synergistic advantage of the LMWG and PG networks. The PG is softened at lower pH while the LMWG benefits of the resistance of the PG network to changes in pH.

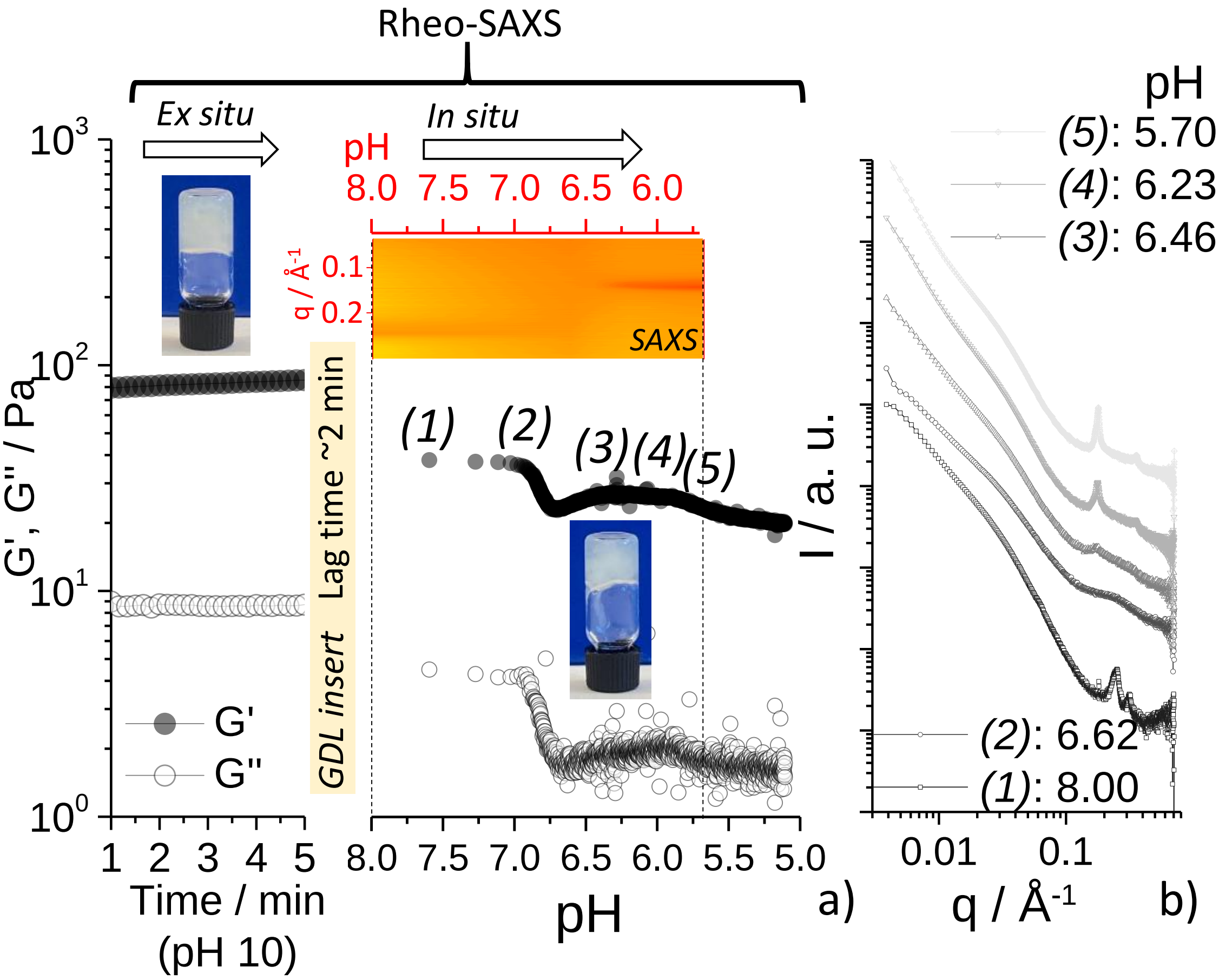

**Figure 4 - GDL-induced acidification of {F}G-C18:1/alginate hydrogel. pH-resolved *in situ* rheo-SAXS experiment: a) mechanical properties (LVER, f= 1 Hz, γ= 0.1 %) and b) corresponding SAXS profiles. The 2D contour plot profile superposed to the G'(pH) evolution in a) corresponds to the full pH-dependent SAXS experiment, synchronized with rheology. $C_{GDL}$= 2 wt%, $C_{G-C18:1}$= 1 wt%, $[Ca^{2+}]$ = 25 mM, $C_{alginate}$= 0.5 wt%. The experimental details of the rheo-SAXS experiment are given in the experimental section. The control rheology experiment performed on alginate at acidic and basic pH is given in Figure S 1b. The lag time in a) is due to the unavoidable security constraints of the SAXS beamline.**

Hybrid {F}G-C18:1/chitosan hydrogels are also submitted to the pH-driven gel-to-sol transition of G-C18:1 and the mechanical properties of the IPN gel are monitored by rheo-SAXS (Figure 5a). When pH is in the order of 8, G' has already lost one order of magnitude, while at pH below 6.4, the gel has lost its properties and forms a viscous solution. This behavior is intermediate between the hybrid gels involving gelatin and alginate, which become viscous or remain a gel, respectively. According to the control experiment in Figure S 1c and to its known solubility at pH below its pKa (6.5), chitosan is expected to lose its elastic properties below pH 6. However, the strong loss in G' between pH 10 and pH 7, above its pKa, also demonstrates the detrimental impact of the gel-to-sol transition of {F}G-C18:1, as found for gelatin and alginate.

The important loss of the elastic properties above the pKa (≈6.5) are explained by the fact that the method employed here to prepare chitosan hydrogels make them highly heterogeneous at basic pH, as they tend to form gel aggregates dispersed in the water phase. The role of the LMWG is then crucial, as it traps chitosan in a homogeneous gel matrix, composed of {F}G-C18:1, as demonstrated by its typical SAXS signature in Figure 5b. All in all, the dynamic self-assembly properties of the LMWG network allow a continuous control of the elastic properties of the chitosan gel in a broad pH range, from pH 6 to 10.

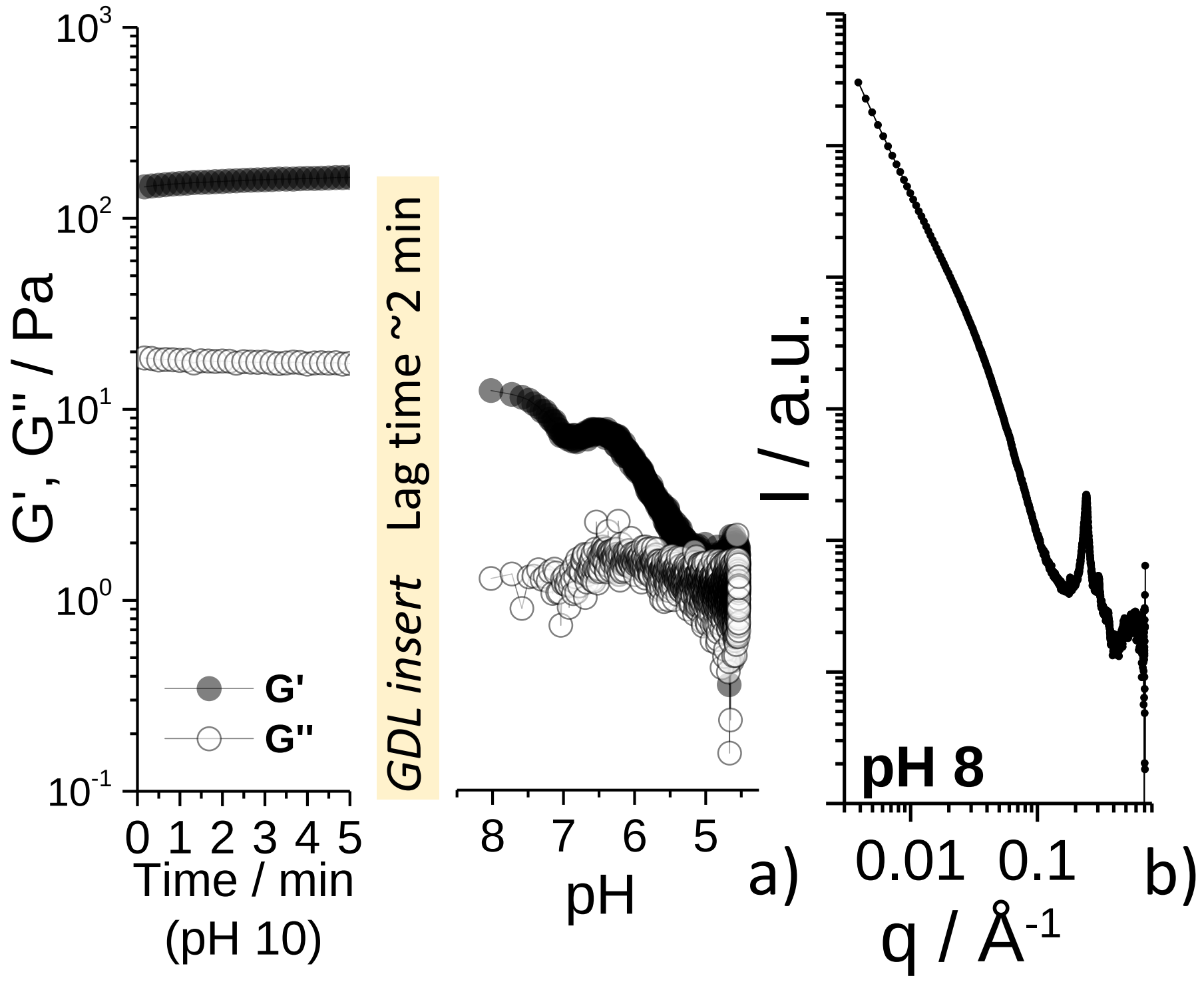

**Figure 5 – a) pH-resolved *in situ* oscillatory rheology (LVER, f= 1 Hz, γ= 0.1 %) of {F}G-C18:1/chitosan HMW using GDL as trigger of acidification. b) SAXS profile of the {F}G-C18:1/chitosan HMW hydrogel at pH 8. Rheo-SAXS experiments could not be adapted to this system. $C_{GDL}$= 10 wt%, $C_{G\text{-}C18:1}$= 1 wt%, $C_{chitosanHMW}$= 0.5 wt %, [$Ca^{2+}$] = 17.8 mM. The control rheology experiment performed on chitosan HMW at acidic and basic pH is given in Figure S 1c. The lag time in a) is due to the unavoidable security constraints of the SAXS beamline.**

*Use of temperature to control the IPN properties*

Temperature is tested as another typical[52–54] stimulus to control the viscoelastic properties of the hybrid hydrogels. Both gelatin and {F}G-C18:1[36] hydrogels undergo a temperature-driven gel-to-sol transition, while alginate and chitosan are less sensitive, as shown by the control rheology experiments presented in Figure S 4. It is then of particular interest to study the double temperature-sensitivity of {F}G-C18:1/gelatin hydrogels but also the effect of temperature on the gel-to-sol transition of {F}G-C18:1 inside the corresponding interpenetrated gels with alginate and chitosan.

The evolution of the storage and loss moduli of {F}G-C18:1, biopolymer and {F}G-C18:1/biopolymer gels was investigated between 20°C and 50°C via a heating-cooling cycle. Figure 6a shows that, as expected, gelatin gel loses its mechanical properties above 28°C and does not reform a gel upon cooling, at least on the experiment's time scale. On the contrary, Figure S 4 illustrates that alginate and chitosan are not sensitive to temperature up to 50°C. Concerning {F}G-C18:1 gels, Figure 6b shows that, although the gel has not reached equilibrium (G' increases after 15 min), the sol-gel transition is promptly reversible on the same time scale compared to gelatin. G' of {F}G-C18:1 is affected starting at about 30°C, culminating with complete loss of the elastic properties between 40°C and 45°C.

The hybrid IPN {F}G-C18:1/gelatin gels are also sensitive to temperature in correspondence of the sol-gel transition of each individual component. Rheo-SAXS of the hybrid {F}G-C18:1/gelatin gel (Figure 6) is performed to associate the elastic and structural properties. Figure 6c shows that the mechanical properties of the hybrid IPN gel initially decrease above 25°C, following the gel-to-sol transition of gelatin (Figure 6a). A second gel-to-sol transition occurs between 30°C and 45°C, precisely corresponding to the gel-to-sol transition of {F}G-C18:1 (Figure 6b). Upon cooling, the elastic properties of {F}G-C18:1/gelatin immediately increase again. This is explained by the reversibility of the sol-gel transition of {F}G-C18:1 rather than gelatin (Figure 6a,b). The contributions of both components are thus nicely established, while the advantage of introducing a fast-responsive self-assembled fibrillary network within a polymer gel now becomes self-evident: the rapid sol-gel transition of {F}G-C18:1 overwhelms the slow recovery of gelatin alone. The fast sol-gel transition of {F}G-C18:1 improves the temperature responsivity of the interpenetrated gel compared to gelatin gel alone.

Figure 6d provides the SAXS profiles associated to the gels at 25°C (a), 50°C (b) and 25°C after cooling (c). The SAXS profiles acquired at different temperatures are characteristics of {F}G-C18:1 and perfectly superimposed, showing that the structure of the gel is not sensitive to temperature at least up to about 50°C. This result is in agreement with previous data, which had shown that the sol-gel transition of {F}G-C18:1, occurring below 50°C, is actually dissociated form its fiber-to-micelle phase transition, found between 60°C and 63°C.[36] This phenomenon is not atypical and it has been reported for other hydrogels prepared from bolaamphiphiles,[55] where authors attributed it to the progressive loss of hydrophobic cross-link interactions across fibers.

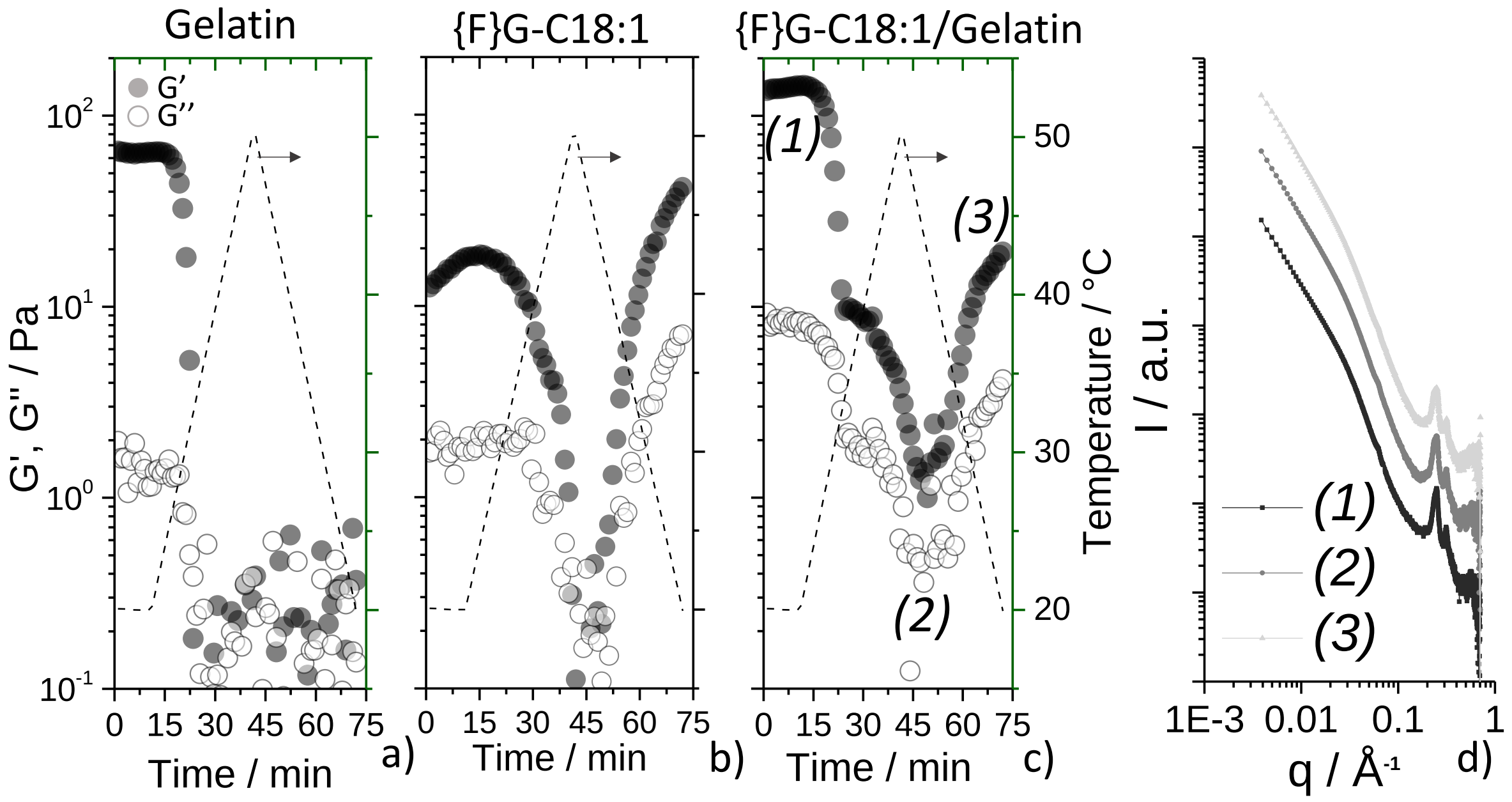

**Figure 6 – Temperature-resolved *in situ* rheo-SAXS experiment (LVER, f= 1 Hz, γ= 0.1 %). Evolution of storage (G') and loss (G'') moduli as a function of temperature of a) gelatin hydrogel, b) {F}G-C18:1 hydrogel and c) hybrid {F}G-C18:1/gelatin gel. $C_{G-C18:1}$= 2 wt%, $C_{gelatin}$= 2 wt%, $[Ca^{2+}]$= 33.5 mM, pH= 8. d) SAXS profiles recorded during the synchronized rheo-SAXS experiments of hybrid {F}G-C18:1/gelatin hydrogels. *(1)* through *(3)* correspond to code numbers in panel c).**

Interpenetrated gels composed of chitosan and alginate were also tested against temperature, knowing that the controls are not temperature-sensitive (Figure S 4). According to the data shown in Figure 7a, the hybrid {F}G-C18:1/alginate gel is only slightly sensitive to temperature, and mainly around 50°C, in correspondence of the gel-to-sol transition of {F}G-C18:1. In this case, the alginate network compensates the weakness of {F}G-C18:1 towards temperature. From a structural point of view, the SAXS profiles acquired at different temperatures (Figure 7b) are identical, thus showing that the gel structure is unaffected by temperature.

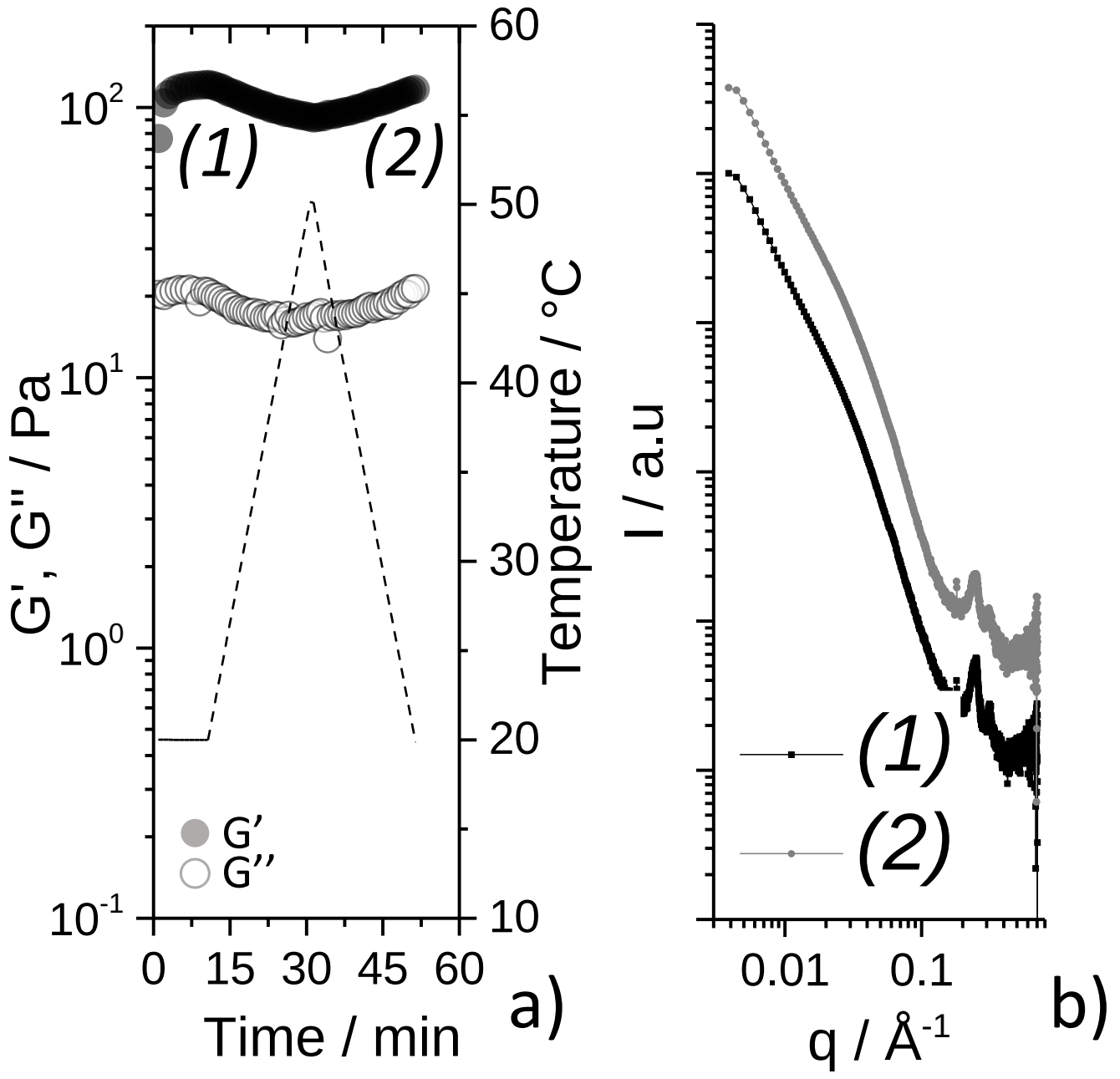

**Figure 7 – Temperature-resolved *in situ* rheo-SAXS experiment (LVER, f= 1 Hz, γ= 0.1 %). Evolution of storage (G') and loss (G'') moduli as a function of temperature of a) hybrid {F}G-C18:1/alginate gel. $C_{G-C18:1}$= 2 wt%, $C_{alginate}$= 1 wt%, [$Ca^{2+}$]= 50 mM, pH= 8. b) SAXS profiles recorded during the synchronized rheo-SAXS experiments of hybrid {F}G-C18:1/alginate hydrogels. *(1)* and *(2)* correspond to code numbers in panel a). The control temperature-resolved rheology experiment performed on alginate is given in Figure S 4a.**

Concerning the properties of hybrid {F}G-C18:1/chitosan gels against temperature, results are quite similar to the ones of alginate : Figure 8a shows that the hybrid {F}G-C18:1/chitosan gel does not have any obvious temperature-responsive property, as also found for the chitosan gel control (Figure S 4b). In this case, chitosan strengthens the hybrid gel compared to the gel of {F}G-C18:1 alone. The SAXS profiles given in Figure 8b demonstrate that, as for the other biopolymers tested, temperature is not a parameter governing the gel structure, at least not up to 50°C.

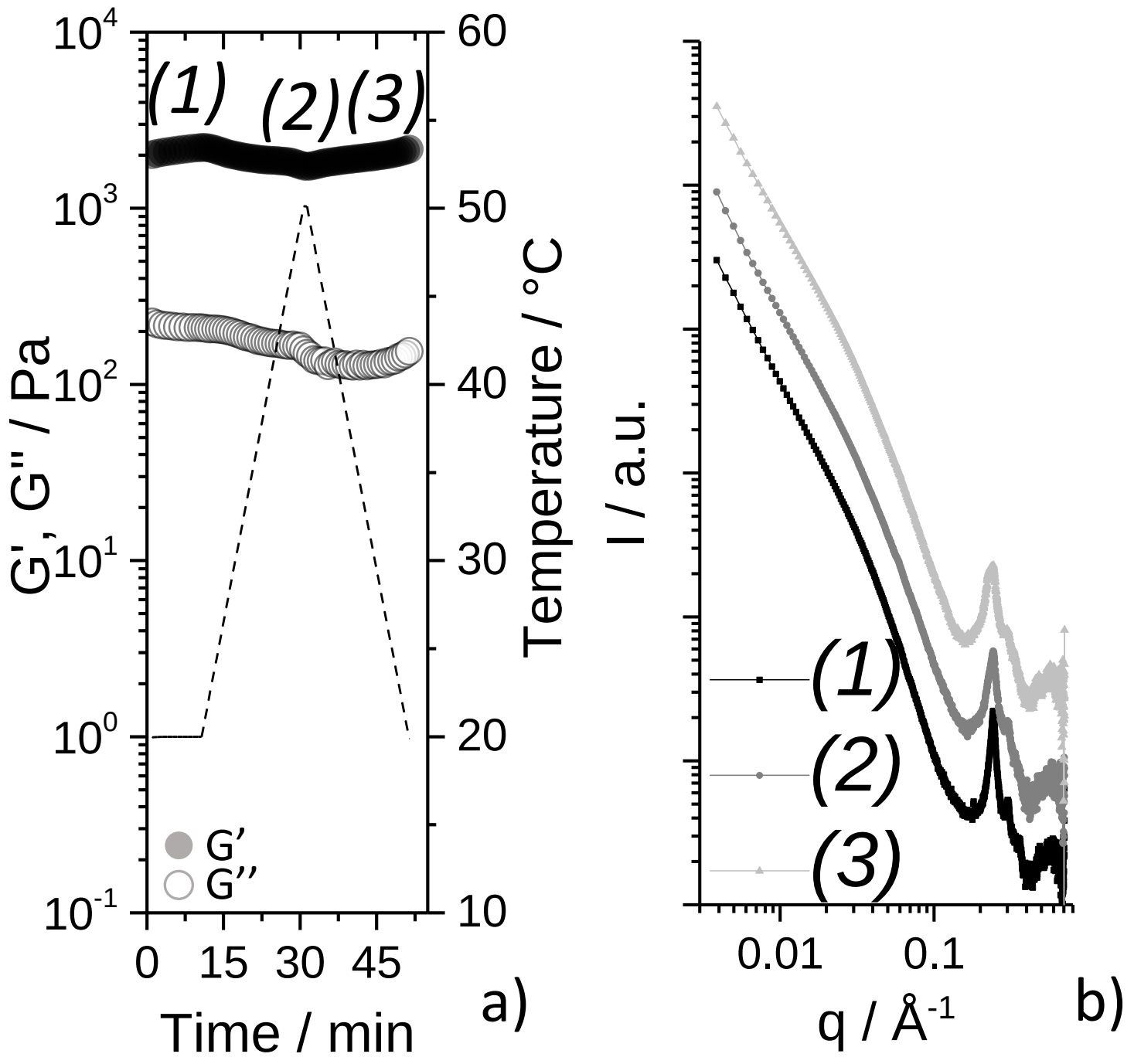

**Figure 8 - Temperature-resolved *in situ* rheo-SAXS experiment (LVER, f= 1 Hz, γ= 0.1 %). Evolution of storage (G') and loss (G'') moduli as a function of temperature of a) hybrid {F}G-C18:1/chitosan HMW gel. $C_{G-C18:1}$= 2 wt%, $C_{chitosanHMW}$= 1 wt%, [$Ca^{2+}$]= 33.5 mM, pH= 8. b) SAXS profiles recorded during the synchronized rheo-SAXS experiments of hybrid {F}G-C18:1/chitosan HMW hydrogels. *(1)* through *(3)* correspond to code numbers in panel a). The control temperature-resolved rheology experiment performed on chitosan HMW is given in Figure S 4b.**

## Discussion

The hybrid {F}G-C18:1/biopolymer hydrogels behave differently towards the applied stimuli and each individual component strongly contributes to improve the overall performances of the IPN. If one of the two gel networks, polymer or LMW, collapses, the second network systematically keeps the system in the gel state (Figure 9a,b). If both networks collapse, the sol-gel transition of the LMW component restores the gel properties of the entire system faster than in the PG control (Figure 9c). Besides these general remarks, which make the present double-responsive PG-LMWG system unique, some additional comments must be formulated.

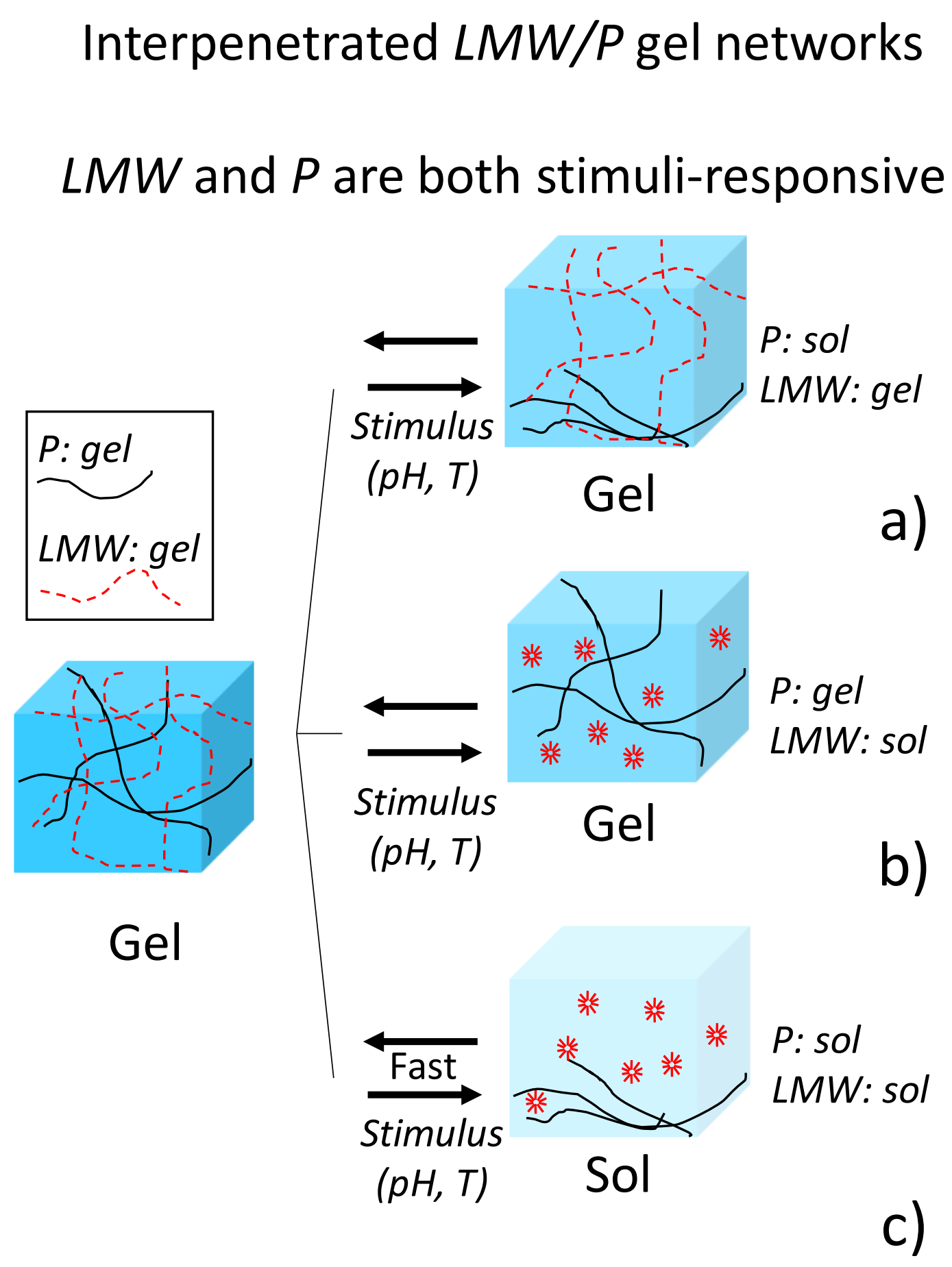

**Figure 9 – Scheme of the advantages of stimuli responsive a low-molecular weight (LMW, segmented line) and polymer (P, solid line) interpenetrated gels. Stimuli are pH and temperature and properties are measured *in situ*.**

*Structural considerations*. A set of morphological (SEM) and spectroscopic (solid-state NMR, ss-NMR, performed under magic angle spinning, MAS) experiments were additionally performed on two representative systems, {F}G-C18:1/gelatin and {F}G-C18:1/alginate, to confirm the rheo-SAXS data and possibly obtain valuable information on the interactions between G-C18:1 and biopolymers. Given the nature of the analytical techniques, involving high vacuum (SEM) or fast MAS at 20 kHz (ss-NMR), the gel state was not adapted. All samples required freeze-drying. Qualitatively-speaking, the freeze-dried samples of gelatin (pH 10 and 4), alginate (pH 10 and 4), {F}G-C18:1 (pH 10) and {F}G-C18:1/biopolymer (alginate, gelatin) at pH 10 appear by the eye as typical, more or less stiff, aerogels. On the other hand, {F}G-C18:1 at pH 4 rehydrate fast and it appears as a highly viscous liquid, while {F}G-C18:1/biopolymer samples, especially those prepared with gelatin, lose the typical aerogel appearance in favor of a more "gummy" look during acidification. The pH values at which acidification was stopped were chosen on the basis of the rheo-SAXS data (Figure 3 and Figure 4), and in particular pH 6.7, where sudden drop of G' occurs, pH 6, where G' has stabilized, and pH 4, where the lamellar phase has appeared.

The SEM images recorded at x500, x2000 and x5000 magnification presented in Figure S 5 show that the acidified {F}G-C18:1/biopolymer samples lose their fibrous structure, always observed at pH 10 and on all controls at both pH 10 and pH 4. The only exception is the control {F}G-C18:1 at pH 4, showing a smooth texture. The progressive fiber loss is enhanced on {F}G-C18:1/gelatin, compared to {F}G-C18:1/alginate, which still shows residual fibers at pH 4. Despite the fact that comparison between hydrated and dehydrated samples should be done with care, due to possible artifacts induced by the freeze-drying process, these results corroborate the idea that the loss in G' during acidification (Figure 3a, Figure 4a) is correlated to a partial collapse of the fibrillar network. The residual elasticity of {F}G-C18:1/alginate (Figure 4a) at pH below 6 could be explained by the residual fibrous structure, shown by SEM in the corresponding freeze-dried samples (Figure S 5). This seems to be most likely attributed to the residual alginate network, less affected by pH than G-C18:1 (controls in Figure S 5). Eventually, SEM corroborates the SAXS data , as it shows that the loss of the elastic properties in {F}G-C18:1/biopolymer can be mostly attributed to the loss in the fibrous structure of G-C18:1 during acidification. Unfortunately, any attempt to exploit SEM data to separate the contribution of either G-C18:1 or biopolymer to the IPN samples was vain.

NMR is a spectroscopic technique, which allows molecular resolution in solution. Employed with a solid-state (ss) magic angle spinning (MAS) probe, it gives access to the spectroscopic features of solid samples.[56] Typical ss-NMR spectra are characterized by broad peaks compared to solution, reflecting the lack of molecular tumbling in the solid state.[57] Complementary $^{1}H$ and $^{13}C$ ss-NMR experiments were then performed on freeze-dried materials in the goal of shedding further light on the possible contributions of either G-C18:1 or biopolymer to the IPN. Results are given in Figure S 6.

Both $^{1}H$ and $^{13}C$ spectra are characterized by the typical signal of aliphatics (~1.2 ppm in $^{1}H$ and ~30 ppm in $^{13}C$), CH in carbohydrates (3-4 ppm in $^{1}H$ and 60-80 ppm in $^{13}C$), C=C (~130 ppm in $^{13}C$) and C=O (~170 ppm in $^{13}C$). If the $^{1}H$ spectra are more difficult to interpret than $^{13}C$ spectra due to their larger line broadening, typical in $^{1}H$ ss-NMR due to strong dipolar coupling, the overall trend is similar between the $^{1}H$ and $^{13}C$ spectra: spectral lines becomes narrower and narrower upon acidification from pH 10 to pH 4 in the G-C18:1/biopolymer samples, for both gelatin and alginate. The $^{13}C$ spectra of the G-C18:1/biopolymer at pH 10 do not show the signals of C=O, C=C and CH (carbohydrate) species (arrows pointing downward in Figure S 6); this is explained by the rigidity of the fibrous network, even with MAS at 20 kHz. On the other hand, from pH 6.7 and below, the peaks corresponding to the same species appear and are narrower during acidification to pH 4 (arrows pointing upward in Figure S 6). Comparison with the control spectra of G-C18:1 and biopolymers (Figure S 6) indicate that line narrowing is solely attributed to the phase change in G-C18:1, from fiber (pH 10, broad spectrum) to lamellar (pH 4, narrow spectrum). Unfortunately, the contribution of the biopolymers is too mild to be resolved in the mixed systems, as shown by comparison with the control $^{13}C$ spectra of gelatin and alginate at pH 10 and pH 4 (Figure S 6), which are practically dominated by the signal of the probe head (broad resonance at 108 ppm, marked by the * symbol).

In summary, ss-NMR confirms that acidification induces structural changes in the mixed G-C18:1/biopolymer systems so that molecular tumbling is more prominent with acidification. This is coherent both with the typical ss-NMR behaviour of lipids[57] and with the known phase change from fiber to wormlike to lamellar of G-C18:1,[25] recalled earlier in the manuscript and discussed elsewhere.[26,27] G-C18:1 in liquid crystalline environment (micelles, membranes) is expected[57] to be more mobile than in crystalline fibers. Unfortunately, similarly to SEM, ss-NMR does not allow to identify neither the contribution of the biopolymer nor the interactions between the biopolymer and G-C18:1 upon acidification.

*The effect of charge*. Cationic polyelectrolytes with high $pK_a$ values strongly interact with G-C18:1 in their micellar and vesicular form, while amphoteric proteins, like gelatin, did not show any remarkable interaction.[30,31] More interestingly, self-assembled glycolipid nanofibers did not show interactions with polyelectrolytes.[31] In the present system, all IPN gels are stable under alkaline pH conditions, when the LMW and biopolymers are like-charged (negative), with the exception of chitosan, which is neutral above pH ≈6.5. The structural SAXS analysis confirms that the LMW fibers and the biopolymer coexist without interacting, similarly to what it was reported before.[31]

Despite the loss of the elastic properties of all IPN gels when pH is reduced, all system are in the gel state at least down to pH ≈6.7-7. Below about pH 6.7, all systems display a sudden drop in G', corresponding to the gel-to-sol transition of G-C18:1. When the pH drops further, each system follows its own trend. {F}G-C18:1/gelatin (Figure 3) loses its elasticity within the time-scale of the experiment. Gelatin is positively-charged (isoelectric point between 7 and 9[33]) and it could interact with G-C18:1. However, the corresponding SAXS data support coexistence of species rather than mutual interactions. This could be most likely due to the increasing neutrality of G-C18:1 at lower pH, as demonstrated by previous ζ-potential expriments.[31]

In the case of {F}G-C18:1/chitosan gels, the gel-to-sol transition of G-C18:1 drives the drop in G' observed between pH 10 and about pH 6.9, similarly to the gelatin and alginate systems. However, G' increases again in the narrow pH gap between 6.9 and 6.4 (Figure 5), while below pH 6.4 G' drops again. This behaviour can be understood by considering the $pK_a$ of chitosan, which is ≈6.5.[58] Above its $pK_a$, chitosan is neutral and in the gel state. This can explain the recovery of the elastic properties below pH 6.9 in the hybrid system. On the other hand, below its $pK_a$, chitosan becomes positively-charged and water-soluble, thus explaining the overall loss in G' of the hybrid gel. If other competing phenomena, like electrostatic attraction between chitosan and G-C18:1 could also occur, they are expected to take place below pH 6.5, thus playing no significant role.

Biosurfactants-biopolymers interactions were studied by us in previous works[30,31] and they were generalized between G-C18:1 and positively-charged polyelectrolytes as follows: when a micellar solution of G-C18:1 is intimately mixed with a low-molecular weight chitosan solution, one observes a phase transition between a complex coacervates (CC) and multilamellar wall vesicles (MLWV), when pH is reduced from 10 to 4.5. This mechanism is unlikely to occur in the present system, because the G-C18:1 phase in {F}G-C18:1/chitosan at alkaline pH is fibrillar and not micellar, thus excluding the CC phase, never detected by SAXS. Secondly, neither the CC nor the MLWV were found to have viscoelastic properties. The rise in G' between pH 6.9 and 6.5 could then not be explained if chitosan would be engaged in an interaction with G-C18:1. Third point, as commented before, the mixing between {F}G-C18:1 and HMW chitosan gels is heterogeneous, thus excluding intimate mixing at the molecular scale. Finally, if electrostatic interactions would still occur and generate a MLWV phase, they would be reasonably observed between pH 6.5 and pH 4.5, where chitosan is positive and G-C18:1 still partially negative and where the MLWV phase was observed before.[30] However, even if this event likely occurs, the rheological properties of the hybrid gel shown in Figure 5 below pH 6.4 would be affected in the same way.

Finally, the {F}G-C18:1/alginate also recovers its elastic properties between pH 6.7 and 6, with G' becoming stable at least down to pH 5. This can be explained by the lower $pK_a$ (≈4) of alginic acid,[34,59] above which complexation with $Ca^{2+}$ ions favors gelation. Below pH ≈4, even the {F}G-C18:1/alginate gel collapses (image taken at pH 3.9 in Figure 2).

In summary, one can identify three different pH-dependent regimes of the G-C18:1/biopolymer systems: 1) in the neutral/alkaline regime with limit at pH 6.7, one can speak of IPN between {F}G-C18:1 and biopolymers, of which the negative, or neutral, state of charge favors gelification. The gel strength probably depends on the pH-controlled fiber-to-micelle ratio.[29,60] 2) The mildly acidic regime, between pH 6.7 and pH 6, is characterized by the gel-to-sol transition of G-C18:1 and a drastic drop in G' occurs in all IPN systems. Overall, the elasticity of the IPN gels in the pH range between 6 and 10 is controlled by the self-assembly of the LMWG. 3) The acidic regime, below pH 6, is controlled by the properties in solution of the biopolymer, which are essentially determined by its state of charge.

**Table 2 – Storage moduli recorded during the pH- (Figure 3, Figure 4, Figure 5, Figure S 1) and temperature-resolved (Figure 6, Figure 7, Figure 8, Figure S 4) *in situ* oscillatory rheology experiments in control (biopolymers) and hybrid ({F}G-C18:1/biopolymer) hydrogels. The variations are given in percentage of loss (-) and gain (+) with respect to the G' values recorded either at pH 8/10 (pH-resolved experiments) or T= 20°C (temperature-resolved).**

| **Sample** | **G' / Pa** | | | | |
|---|---|---|---|---|---|
| | pH-resolved | | | Temperature-resolved | |
| | pH 8/10 | pH 6 | pH 5 | T= 20°C | T= 50°C |
| gelatin | 49 | - | 27 / -45% | 49 | G'<G'' |
| {F}G-C18:1/gelatin | 100 | G'<G'' | G'<G'' | 140 | 1.2 / -99% |
| alginate | 124 | 74 / -40% | - | 38 | 37 / -3% |
| {F}G-C18:1/alginate | 84 | 21 / -75% | 18 / -79% | 122 | 89 / -27% |
| chitosan | 420 | 105 / -75% | - | 44 | 46 / +5% |
| {F}G-C18:1/chitosan | 161 | 5.8 / -96% | G'<G'' | 2890 | 1913 / -34% |

*The effect of kinetics*. The acidification kinetics seems to play an important impact on the elastic properties in the entire pH range, with particular effects in the neutral/alkaline pH regime. The difference in terms of the mechanical properties of a IPN hydrogel independently prepared at pH 8 and pH 6 is only of few tens of Pascal.[23] The difference is much more pronounced using *in situ*, GDL-triggered, acidification. In the neutral/alklaine regime, G' drops of tens of Pa in all systems. More specifically, {F}G-C18:1/alginate loses over 70% of its G' (-75% at pH 6 and -79% at pH 5, Figure 4a, Table 2). For {F}G-C18:1/chitosan, the loss in G' is even higher already at pH 7, it reaches -96% at pH 6 and catastrophic at pH 5, where a gel no longer exists (G'< G'') (Figure 5a, Table 2). In the case of gelatin, elastic properties (G'< G'') are entirely lost already at pH 7 (Figure 3a).

To explain the discrepancy between the elastic properties of IPN systems measured at[23] and out of (this work) equilibrium, one must consider a series of factors. First of all, {F}G-C18:1 gels have higher G' at pH 10 than at pH 8,[36] probably explained by a different fiber-to-micelle ratio between these pH values. However, this evidence does not explain the discrepancies in the magnitude of the drops, ranging from few tens of Pascal ({F}G-C18:1/alginate) to a decade ({F}G-C18:1/chitosan). The discrepancy could probably be explained by kinetics effects, where the profile of pH drops generally found with GDL is not linear, but consists in an initial drastic pH drop, followed by progressive stabilization over time.[38,39,42] Considering that hydrolysis of GDL is hard to control,[61] and considering that small variations in the ratio with other acids and bases can strongly affect the material's properties, as seen before for similar systems,[38] one can reasonably suppose that the necessarily uncontrolled initial pH drop could have a different impact on the gel's strength from one system to another. GDL has another intrinsic problem, it is a powder and it has to be added as such. In this regard, even if all solutions are systematically vortexed before measurement, one should not underestimate their gel state. Homogenization of GDL could not be even across samples. An additional problem could also come from the concomitant acido-base reaction of the biopolymer, each having a different $pK_a$, and competing with G-C18:1. Finally, it is not excluded that kinetics factors related to the diffusivity of hydronium ions generate a mismatch between the solution pH and the actual state of protonation of each individual acid and base. Similar effects have been observed before.[38]

*Interactions*. Differently than in other systems, where strong interactions between amphiphiles and polyelectrolytes generally modify the scattering pattern of their mixture,[30,31,62,63] the structural SAXS data recorded on IPN gels throughout this work always show the typical signature of the {F}G-C18:1 control. Within the limit of the present q-range, SAXS data do not show any influence of the self-assembled fibrillar hydrogel on the mesh size of the biopolymer.[23] On the contrary, variations in the slope of the Ornstein-Zernike plots associated to the biopolymer hydrogel control and biopolymer inside the hybrid network suggested a qualitative relationship between the micellar phase and the mesh size of the biopolymer, as found for other systems in the literature.[64–66] According to the above, it is not unreasonable to think that the fiber-to-micelle transition of G-C18:1 in the neutral/acidic regimes affects the mesh size of the biopolymer network in the {F}G-C18:1/gelatin system, but not for {F}G-C18:1/alginate and {F}G-C18:1/chitosan, of which the elasticity increases again in the mildly acidic regime, between pH 6.7 and pH 6. Specific interactions between G-C18:1 and gelatin are of course not excluded, as observed for anionic surfactants.[67] In fact, interactions between anionic glycolipid biosurfactants and proteins have also been reported before.[68] However, the experimental techniques involved in this study do allow to confirm the presence of such interactions.

*Effect of temperature*. The effect of temperature is mainly observed in the gelatin-based hybrid gels. The loss in the elastic properties occurs in a two-step process. The first sol-gel transition is driven by gelatin at about 28°C (Figure 6a,c), while the second one above 30°C is driven by G-C18:1 (Figure 6b,c). Interestingly, the constant signal of the associated SAXS data tells that the loss in the elastic properties is explained by the enhanced fluidity of the fiber phase, rather than by a fiber-to-micelle phase transition of G-C18:1, the latter reported for temperatures above 60°C.[36] The most striking contribution of {F}G-C18:1 to the gelatin gel is undoubtedly the rapid increase in the elastic properties upon cooling, attributed to the sol-to-gel transition in G-C18:1 (Figure 6b) and illustrating the importance of the LMWG/PG interpenetrated network (Figure 9a,c).

Alginate and chitosan hydrogels are practically not sensitive to temperature if compared to gelatin. The temperature-dependent evolution of G' for the controls shows a bare loss of -3% for alginate and an actual improvement of +5% for chitosan (Figure S 4, Table 2). Such a stability is also found in the hybrid {F}C18:1/alginate and {F}G-C18:1/chitosan gels, which are obviously less sensitive to temperature if compared to the hybrid {F}C18:1/gelatin hydrogel. Nonetheless, the presence of the {F}G-C18:1 fiber phase in the alginate and chitosan network still has a non-negligible influence on the elastic properties. The loss in G' between room temperature and 50°C is of -27% and -34% for {F}C18:1/alginate (Figure 7a, Table 2) and {F}G-C18:1/chitosan (Figure 8a, Table 2), respectively, whereas the gel structure does not vary, as shown by the corresponding SAXS profiles (Figure 7b, Figure 8b). The data collected on the hybrid alginate and chitosan gels are also of particular interest, as they show the strong interest of the hybrid network compared to the individual components: the presence of {F}G-C18:1 introduces a modest, although measurable, responsiveness to temperature. Considering the stiffness of the control gels, temperature could then be used as a way to control their rheological properties and make them easier to process. On the other hand, the biopolymer network gives a much better stability to the {F}G-C18:1 hydrogels, compared to the SAFIN gel alone (Figure 9b).

## Conclusion

This work shows the stimuli responsivity of an interpenetrated hydrogel network composed of a bioamphiphile and a biopolymer. The bioamphiphile is a microbially-produced glycolipid, able to self-assemble, in water, in a variety of structures, fibers, micelles or vesicles, according to the type of ion and pH. Here, a fiber phase of G-C18:1 ({F}G-C18:1) with independent hydrogelating properties is mixed with biopolymer hydrogel (gelatin, chitosan or alginate). SAXS and rheology show that the self-assembled fibrillar network of G-C18:1 is interpenetrated with the biopolymer's network and the overall elastic properties are improved, if compared to the properties of each individual system. The use of pH as a stimulus, here by mean of a cyclic lactone, GDL, induces a prompt collapse of the interpenetrated {F}G-C18:1/gelatin and {F}G-C18:1/chitosan hydrogels, whereas {F}G-C18:1/alginate hydrogels experience a loss in G', as well, but of minor importance. Comparison with the pH-dependent properties of each biopolymer and observations in the structure of G-C18:1 by SAXS show that the loss in G' is mainly related to the pH-dependent phase transition of G-C18:1 itself inside the interpenetrated gels. This is confirmed by complementary SEM and solid-state $^{1}H$ and $^{13}C$ MAS NMR experiments.

Effect of temperature on the interpenetrated gels also varies across samples. For the {F}G-C18:1/gelatin, temperature individually affects the elasticity of gelatin (28°C) and {F}G-C18:1 (>30°C). However, the faster sol-gel transition of {F}G-C18:1 compared to gelatin shows that the properties of the interpenetrated gel can be promptly recovered within minutes, instead of hours in the case of gelatin alone. This behavior is directly associated to the {F}G-C18:1 component. On the other hand, the elasticity of alginate and chitosan gels are much less sensitive to temperature, compared to {F}G-C18:1. In the corresponding interpenetrated gels, one observes that the elastic moduli are affected in the order of 30 ± 5 % when temperature is increased to 50°C. Differently than in the gelatin case, the use of chitosan or alginate shows the importance of mixing these biopolymers with {F}G-C18:1: the gel-to-sol transition in {F}G-C18:1 slightly reduces the elastic properties of the biopolymers, thus favoring their manipulation, for instance. On the other hand, the stability of the biopolymers' networks to temperature strongly compensates, and stabilizes, the collapse of the {F}G-C18:1 gels.

**Acknowledgements**

Soleil Synchrotron facility is acknowledged for accessing the Swing beamline and financial support. Ghazi Ben Messaoud (DWI-Leibniz Institute for Interactive Materials, Aachen, Germany) is kindly acknowledged for helpful discussions. We thank Dr. S. Roelants and Prof. W. Soetaert at Gent University for shipping to us the glycolipid. Sorbonne Université (contract N° 3083/2018) is acknowledged for financial support of CS. Authors kindly acknowledge the French ANR, Project N° SELFAMPHI - 19-CE43-0012-01. Isabelle Genois (Sorbonne Université, Paris, France) is kindly acknowledged for her assistance on SEM experiments. Christina Coelho (Institut des Matériaux de Paris Centre, Sorbonne Université, Paris, France) is kindly acknowledged for her assistance on the ss-NMR experiments.

**Supporting Information**

Time-dependence of G' and G'' (Figure S 1), comparison between the SAXS profiles of {F}G-C18:1/Gelatin, {F}G-C18:1/Alginate and G-C18:1 (Figure S 2), low-q scaling of the SAXS scattered intensity for G-C18:1 and {F}G-C18:1/biopolymer (Figure S 3), temperature dependence of G' and G'' (Figure S 4), SEM (Figure S 5) and $^{1}H$, $^{13}C$ solid-state MAS NMR (Figure S 6) experiments collected for freeze-dried hydrogels prepared from: {F}G-C18:1, gelatin, alginate, {F}G-C18:1/biopolymer. This work benefited from the use of the SasView application, originally developed under NSF award DMR-0520547. SasView contains code developed with funding from the European Union's Horizon 2020 research and innovation programme under the SINE2020 project, grant agreement No 654000

**References**

(1) Dhand, A. P.; Galarraga, J. H.; Burdick, J. A. Enhancing Biopolymer Hydrogel Functionality through Interpenetrating Networks. *Trends Biotechnol.* **2021**, *39*, 519–538.

(2) Zhang, Y. S.; Khademhosseini, A. Advances in Engineering Hydrogels. *Science (80-. ).* **2017**, *356*, 3627.

(3) Mohammadzadeh Pakdel, P.; Peighambardoust, S. J. Review on Recent Progress in Chitosan-Based Hydrogels for Wastewater Treatment Application. *Carbohydr. Polym.* **2018**, *201*, 264–279.

(4) Gholizadeh, H.; Cheng, S.; Pozzoli, M.; Messerotti, E.; Traini, D.; Young, P.; Kourmatzis, A.; Ong, H. X. Smart Thermosensitive Chitosan Hydrogel for Nasal Delivery of Ibuprofen to Treat Neurological Disorders. *Expert Opin. Drug Deliv.* **2019**, *16*, 453–466.

(5) Wu, B. C.; Degner, B.; McClements, D. J. Soft Matter Strategies for Controlling Food Texture: Formation of Hydrogel Particles by Biopolymer Complex Coacervation. *J. Phys. Condens. Matter* **2014**, *26*, 464104.

(6) Panja, S.; Adams, D. J. Stimuli Responsive Dynamic Transformations in Supramolecular Gels. *Chem. Soc. Rev.* **2021**, *50*, 5165–5200.

(7) Cornwell, D. J.; Smith, D. K. Expanding the Scope of Gels – Combining Polymers with Low-Molecular-Weight Gelators to Yield Modified Self-Assembling Smart Materials with High-Tech Applications. *Mater. Horiz.* **2015**, *2*, 279–293.

(8) Stubenrauch, C.; Gießelmann, F. Gelled Complex Fluids: Combining Unique

Structures with Mechanical Stability. *Angew. Chemie - Int. Ed.* **2016**, *55*, 3268–3275.

(9) Poolman, J. M.; Boekhoven, J.; Besselink, A.; Olive, A. G. L.; Van Esch, J. H.; Eelkema, R. Variable Gelation Time and Stiffness of Low-Molecular-Weight Hydrogels through Catalytic Control over Self-Assembly. *Nat. Protoc.* **2014**, *9*, 977–988.

(10) Koehler, J. PH-Modulating Poly(Ethylene Glycol)/Alginate Hydrogel Dressings for the Treatment of Chronic Wounds. *Macromol. Biosci.* **2017**, *17*, Epub 2016.

(11) Panja, S.; Dietrich, B.; Shebanova, O.; Smith, A. J.; Adams, D. J. Programming Gels Over a Wide PH Range Using Multicomponent Systems. *Angew. Chemie - Int. Ed.* **2021**, *60*, 9973–9977.

(12) Panja, S.; Seddon, A.; Adams, D. J. Controlling Hydrogel Properties by Tuning Non-Covalent Interactions in a Charge Complementary Multicomponent System. *Chem. Sci.* **2021**, *12*, 11197–11203.

(13) Piras, C. C.; Smith, D. K. Sequential Assembly of Mutually Interactive Supramolecular Hydrogels and Fabrication of Multi-Domain Materials. *Chem. - A Eur. J.* **2019**, *25*, 11318–11326.

(14) Cornwell, D. J.; Okesola, B. O.; Smith, D. K. Hybrid Polymer and Low Molecular Weight Gels-Dynamic Two-Component Soft Materials with Both Responsive and Robust Nanoscale Networks. *Soft Matter* **2013**, *9*, 8730–8736.

(15) Cornwell, D. J.; Okesola, B. O.; Smith, D. K. Multidomain Hybrid Hydrogels: Spatially Resolved Photopatterned Synthetic Nanomaterials Combining Polymer and Low-Molecular-Weight Gelators. *Angew. Chemie - Int. Ed.* **2014**, *53*, 12461–12465.

(16) Cornwell, D. J.; Smith, D. K. Photo-Patterned Multi-Domain Multi-Component Hybrid Hydrogels. *Chem. Commun.* **2020**, *56*, 7029–7032.

(17) Cornwell, D. J.; Daubney, O. J.; Smith, D. K. Photopatterned Multidomain Gels: Multi-Component Self-Assembled Hydrogels Based on Partially Self-Sorting 1,3:2,4-Dibenzylidene- d -Sorbitol Derivatives. *J. Am. Chem. Soc.* **2015**, *137*, 15486–15492.

(18) Kim, A. R.; Lee, S. L.; Park, S. N. Properties and in Vitro Drug Release of PH- and Temperature-Sensitive Double Cross-Linked Interpenetrating Polymer Network Hydrogels Based on Hyaluronic Acid/Poly (N-Isopropylacrylamide) for Transdermal Delivery of Luteolin. *Int. J. Biol. Macromol.* **2018**, *118*, 731–740.

(19) Zhao, J.; Zhao, X.; Guo, B.; Ma, P. X. Multifunctional Interpenetrating Polymer Network Hydrogels Based on Methacrylated Alginate for the Delivery of Small Molecule Drugs and Sustained Release of Protein. *Biomacromolecules* **2014**, *15*, 3246–

3252.
(20) Chakraborty, P.; Roy, B.; Bairi, P.; Nandi, A. K. Improved Mechanical and Photophysical Properties of Chitosan Incorporated Folic Acid Gel Possessing the Characteristics of Dye and Metal Ion Absorption. *J. Mater. Chem* **2012**, *22*, 20291–20298.
(21) Chen, L.; Revel, S.; Morris, K.; Spiller, D. G.; Serpell, L. C.; Adams, D. J. Low Molecular Weight Gelator–Dextran Composites. *Chem. Commun.* **2010**, *46*, 6738–6740.
(22) Yang, C.; Bian, M.; Yang, Z. A Polymer Additive Boosts the Anti-Cancer Efficacy of Supramolecular Nanofibers of Taxol. *Biomater. Sci.* **2014**, *2*, 651–654.
(23) Seyrig, C.; Poirier, A.; Perez, J.; Bizien, T.; Baccile, N. Interpenetrated Biosurfactant-Biopolymer Orthogonal Hydrogels: The Biosurfactant's Phase Controls the Hydrogel's Mechanics. *Biomacromolecules* **2022**, Just Accepted, DOI: 10.1021/acs.biomac.2c00319.
(24) Dowling, M. B.; Lee, J.-H.; Raghavan, S. R. PH-Responsive Jello: Gelatin Gels Containing Fatty Acid Vesicles. *Langmuir* **2009**, *25*, 8519–8525.
(25) Baccile, N.; Poirier, A.; Seyrig, C.; Griel, P. Le; Perez, J.; Hermida-Merino, D.; Pernot, P.; Roelants, S.; Soetaert, W. Chameleonic Amphiphile: The Unique Multiple Self-Assembly Properties of a Natural Glycolipid in Excess of Water. *J. Colloid Interface Sci.* **2023**, *630*, 404–415.
(26) Baccile, N.; Cuvier, A.-S.; Prévost, S.; Stevens, C. V; Delbeke, E.; Berton, J.; Soetaert, W.; Van Bogaert, I. N. A.; Roelants, S. Self-Assembly Mechanism of PH-Responsive Glycolipids: Micelles, Fibers, Vesicles, and Bilayers. *Langmuir* **2016**, *32*, 10881–10894.
(27) Baccile, N.; Selmane, M.; Le Griel, P.; Prévost, S.; Perez, J.; Stevens, C. V.; Delbeke, E.; Zibek, S.; Guenther, M.; Soetaert, W.; et al. PH-Driven Self-Assembly of Acidic Microbial Glycolipids. *Langmuir* **2016**, *32*, 6343–6359.
(28) Poirier, A.; Griel, P. Le; Perez, J.; Hermida-Merino, D.; Pernot, P.; Baccile, N. Metallogels from Glycolipid Biosurfactant. *ACS Sustain. Chem. Eng.* **2022**, Just Accepted, DOI: 10.1021/acssuschemeng.2c01860.
(29) Poirier, A.; Griel, P. Le; Hoffmann, I.; Perez, J.; Pernot, P.; Fresnais, J.; Baccile, N. Ca2+ and Ag+ Orient Low-Molecular Weight Amphiphile Self-Assembly into "Nano-Fishnet" Fibrillar Hydrogels with Unusual β-Sheet-like Raft Domains. *Soft Matter* **2022**, Just Accepted, DOI: 10.1039/D2SM01218A.

(30) Seyrig, C.; Griel, P. Le; Cowieson, N.; Perez, J.; Baccile, N. Synthesis of Multilamellar Walls Vesicles Polyelectrolyte-Surfactant Complexes from PH-Stimulated Phase Transition Using Microbial Biosurfactants. *J. Colloid Interface Sci.* **2020**, *580,* 493–502.

(31) Seyrig, C.; Kignelman, G.; Thielemans, W.; Griel, P. Le; Cowieson, N.; Perez, J.; Baccile, N. Stimuli-Induced Non-Equilibrium Phase Transitions in Polyelectrolyte-Surfactant Complex Coacervates. *Langmuir* **2020**, *36*, 8839–8857.

(32) Holmberg, K.; Jönsson, B.; Kronberg, B.; Lindman, B. Surfactant-Polymer Systems. In *Surfactants and Polymers in Aqueous Solution*; John Wiley & Sons: Chichester, WestSussex, England, 2002; pp 277–303.

(33) Djabourov, M. Architecture of Gelatin Gels. *Contemp. Phys.* **1988**, *29*, 273.

(34) Martinsen, A.; Skjaak-Braek, G.; Smidsroed, O. Alginate as Immobilization Material : I. Correlation between Chemcial and Physical Properties of Alginate Gel Beads. *Biotechnol. Bioeng.* **1989**, *33*, 79–89.

(35) Rowley, J. A.; Madlambayan, G.; Mooney, D. J. Alginate Hydrogels as Synthetic Extracellular Matrix Materials. *Biomaterials* **1999**, *20*, 45–53.

(36) Poirier, A.; Bizien, T.; Zinn, T.; Pernot, P.; Baccile, N. Shear Recovery and Temperature Stability of Ca2+ and Ag+ Glycolipid Fibrillar Metallogels with Unusual β-Sheet-like Domains. *Soft Matter* **2022**, Just Accepted, DOI: 10.1039/d2sm00374k.

(37) Saerens, K. M. J.; Zhang, J.; Saey, L.; Van Bogaert, I. N. A.; Soetaert, W. Cloning and Functional Characterization of the UDP-Glucosyltransferase UgtB1 Involved in Sophorolipid Production by Candida Bombicola and Creation of a Glucolipid-Producing Yeast Strain. *Yeast* **2011**, *28*, 279–292.

(38) Ben Messaoud, G.; Le Griel, P.; Hermida-Merino, D.; Roelants, S. L. K. W.; Soetaert, W.; Stevens, C. V.; Baccile, N. PH-Controlled Self-Assembled Fibrillar Network (SAFiN) Hydrogels: Evidence of a Kinetic Control of the Mechanical Properties. *Chem. Mater.* **2019**, *31*, 4817–4830.

(39) Adams, D. J.; Butler, M. F.; Frith, W. J.; Kirkland, M.; Mullen, L.; Sanderson, P. A New Method for Maintaining Homogeneity during Liquid–Hydrogel Transitions Using Low Molecular Weight Hydrogelators. *Soft Matter* **2009**, *5*, 1856.

(40) Adams, D. J.; Mullen, L. M.; Berta, M.; Chen, L.; Frith, W. J. Relationship between Molecular Structure, Gelation Behaviour and Gel Properties of Fmoc-Dipeptides. *Soft Matter* **2010**, *6*, 1971–1980.

(41) Raeburn, J.; Pont, G.; Chen, L.; Cesbron, Y.; Lévy, R.; Adams, D. J. Fmoc-

Diphenylalanine Hydrogels: Understanding the Variability in Reported Mechanical Properties. *Soft Matter* **2012**, *8*, 1168–1174.

(42) Chen, L.; Morris, K.; Laybourn, A.; Elias, D.; Hicks, M. R.; Rodger, A.; Serpell, L.; Adams, D. J. Self-Assembly Mechanism for a Naphthalene-Dipeptide Leading to Hydrogelation. *Langmuir* **2010**, *26*, 5232–5242.

(43) Ben Messaoud, G.; Griel, P. Le; Merino, D. H.; Baccile, N. Effect of PH, Temperature and Shear on the Structure-Property Relationship of Lamellar Hydrogels from Microbial Glycolipid Probed by in-Situ Rheo-SAXS. *Soft Matter* **2020**, *16*, 2540–2551.

(44) Piras, C. C.; Mahon, C. S.; Genever, P. G.; Smith, D. K. Shaping and Patterning Supramolecular Materials Stem Cell-Compatible Dual-Network Hybrid Gels Loaded with Silver Nanoparticles. *ACS Biomater. Sci. Eng.* **2022**, *8*, 1829−1840.

(45) Gurikov, P.; Smirnova, I. Non-Conventional Methods for Gelation of Alginate. *Gels* **2018**, *4*, 14.

(46) SasView. SasView 3.1.2 Documentation https://www.sasview.org/docs/old_docs/3.1.2/index.html (accessed Nov 15, 2022).

(47) Glatter, O.; Kratky, O. *Small Angle X-Ray Scattering*; Academic Press: London, 1982.

(48) Stradner, A.; Glatter, O.; Schurtenberger, P. Hexanol-Induced Sphere-to-Flexible Cylinder Transition in Aqueous Alkyl Polyglucoside Solutions. *Langmuir* **2000**, *16*, 5354–5364.

(49) Zhang, R.; Marone, P. A.; Thiyagarajan, P.; Tiede, D. M. Structure and Molecular Fluctuations of N-Alkyl-β-D-Glucopyranoside Micelles Determined by X-Ray and Neutron Scattering. *Langmuir* **1999**, *15*, 7510–7519.

(50) Wei, Y.; Hore, M. J. A. Characterizing Polymer Structure with Small-Angle Neutron Scattering: A Tutorial. *J. Appl. Phys.* **2021**, *129*, 171101.

(51) Pedersen, J. S.; Schurtenbercer, P. Scattering Functions of Semidilute Solutions of Polymers in a Good Solvent. *J. Polym. Sci. Part B Polym. Phys.* **2004**, *42*, 3081–3094.

(52) Yin, X.; Hoffman, A. S.; Stayton, P. S. Poly(N-Isopropylacrylamide-Co-Propylacrylic Acid) Copolymers That Respond Sharply to Temperature and PH. *Biomacromolecules* **2006**, *7*, 1381–1385.

(53) Nakayama, M.; Okano, T.; Miyazaki, T.; Kohori, F.; Sakai, K.; Yokoyama, M. Molecular Design of Biodegradable Polymeric Micelles for Temperature-Responsive Drug Release. *J. Control. Release* **2006**, *115*, 46–56.

(54) Coughlan, D. C.; Quilty, F. P.; Corrigan, O. I. Effect of Drug Physicochemical Properties on Swelling/Deswelling Kinetics and Pulsatile Drug Release from Thermoresponsive Poly(N-Isopropylacrylamide) Hydrogels. *J. Control. Release* **2004**, *98*, 97–114.

(55) Meister, A.; Bastrop, M.; Koschoreck, S.; Garamus, V. M.; Sinemus, T.; Hempel, G.; Drescher, S.; Dobner, B.; Richtering, W.; Huber, K.; et al. Structure-Property Relationship in Stimulus-Responsive Bolaamphiphile Hydrogels. *Langmuir* **2007**, *23*, 7715–7723.

(56) Mehring, M. *Principles of High Resolution NMR in Solids*; Springer London, Limited, 2012.

(57) Grelard, A.; Couvreux, A.; Loudet, C.; Dufourc, E. J. Lipid Signaling Protocols: Solution and Solid-State NMR of Lipids. In *Lipid Signaling Protocols*; Waugh, M. G., Ed.; Humana Press: Totowa, NJ, 2009; Vol. 462, pp 111–133.

(58) Zargar, V.; Asghari, M.; Dashti, A. A Review on Chitin and Chitosan Polymers : Structures, Chemistry, Solubility, Derivatives, and Applications. *Chem. Bio. Eng. Rev.* **2015**, *2*, 204–226.

(59) Fang, Y.; Al-Assaf, S.; Phillips, G. O.; Nishinari, K.; Funami, T.; Williams, P. A.; Li, A. Multiple Steps and Critical Behaviors of the Binding of Calcium to Alginate. *J. Phys. Chem. B* **2007**, *111*, 2456–2462.

(60) Poirier, A.; Griel, P. Le; Perez, J.; Baccile, N. Cation-Induced Fibrillation of Microbial Glycolipid Biosurfactant Probed by Ion-Resolved In Situ SAXS. *J. Phys. Chem. B* **2022**, Just Accepted, DOI: 10.1021/acs.jpcb.2c03739.

(61) Pocker, Y.; Green, E. Hydrolysis of D-Glucono-δ-Lactone. I. General Acid—Base Catalysis, Solvent Deuterium Isotope Effects, and Transition State Characterization. *J. Am. Chem. Soc.* **1973**, *95*, 113–119.

(62) Chiappisi, L.; David Leach, S.; Gradzielski, M. Precipitating Polyelectrolyte-Surfactant Systems by Admixing a Nonionic Surfactant-a Case of Cononsurfactancy. *Soft Matter* **2017**, *13*, 4988–4996.

(63) Chiappisi, L.; Prévost, S.; Grillo, I.; Gradzielski, M. From Crab Shells to Smart Systems: Chitosan-Alkylethoxy Carboxylate Complexes. *Langmuir* **2014**, *30*, 10615–10616.

(64) Pekař, M. Hydrogels with Micellar Hydrophobic (Nano)Domains. *Front. Mater.* **2015**, *1*, 1–14.

(65) Wallace, M.; Adams, D. J.; Iggo, J. A. Analysis of the Mesh Size in a Supramolecular Hydrogel by PFG-NMR Spectroscopy. *Soft Matter* **2013**, *9*, 5483–5491.

(66) Zustiak, S. P.; Leach, J. B. Hydrolytically Degradable Poly(Ethylene Glycol) Hydrogel

Scaffolds with Tunable Degradation and Mechanical Properties Silviya. *Biomacromolecules* **2010**, *11*, 1348–1357.

(67) Greener, J.; Contestable, B. A.; Bale, M. D. Interaction of Anionic Surfactants with Gelatin: Viscosity Effects. *Macromolecules* **1987**, *20*, 2490–2498.

(68) Otzen, D. Protein-Surfactant Interactions: A Tale of Many States. *Biochim. Biophys. Acta - Proteins Proteomics* **2011**, *1814*, 562–591.

## Supplementary Information

# *In situ* stimulation of self-assembly tunes the elastic properties of interpenetrated biosurfactant-biopolymer hydrogels

Chloé Seyrig,[a] Alexandre Poirier,[a] Thomas Bizien,[b] Niki Baccile[a]

[a] Sorbonne Université, Centre National de la Recherche Scientifique, Laboratoire de Chimie de la Matière Condensée de Paris , LCMCP, F-75005 Paris, France

[b] Synchrotron SOLEIL, L'Orme des Merisiers Saint-Aubin, BP 48 91192 Gif-sur-Yvette Cedex

* Corresponding author:
Dr. Niki Baccile
E-mail address: niki.baccile@sorbonne-universite.fr
Phone: +33 1 44 27 56 77

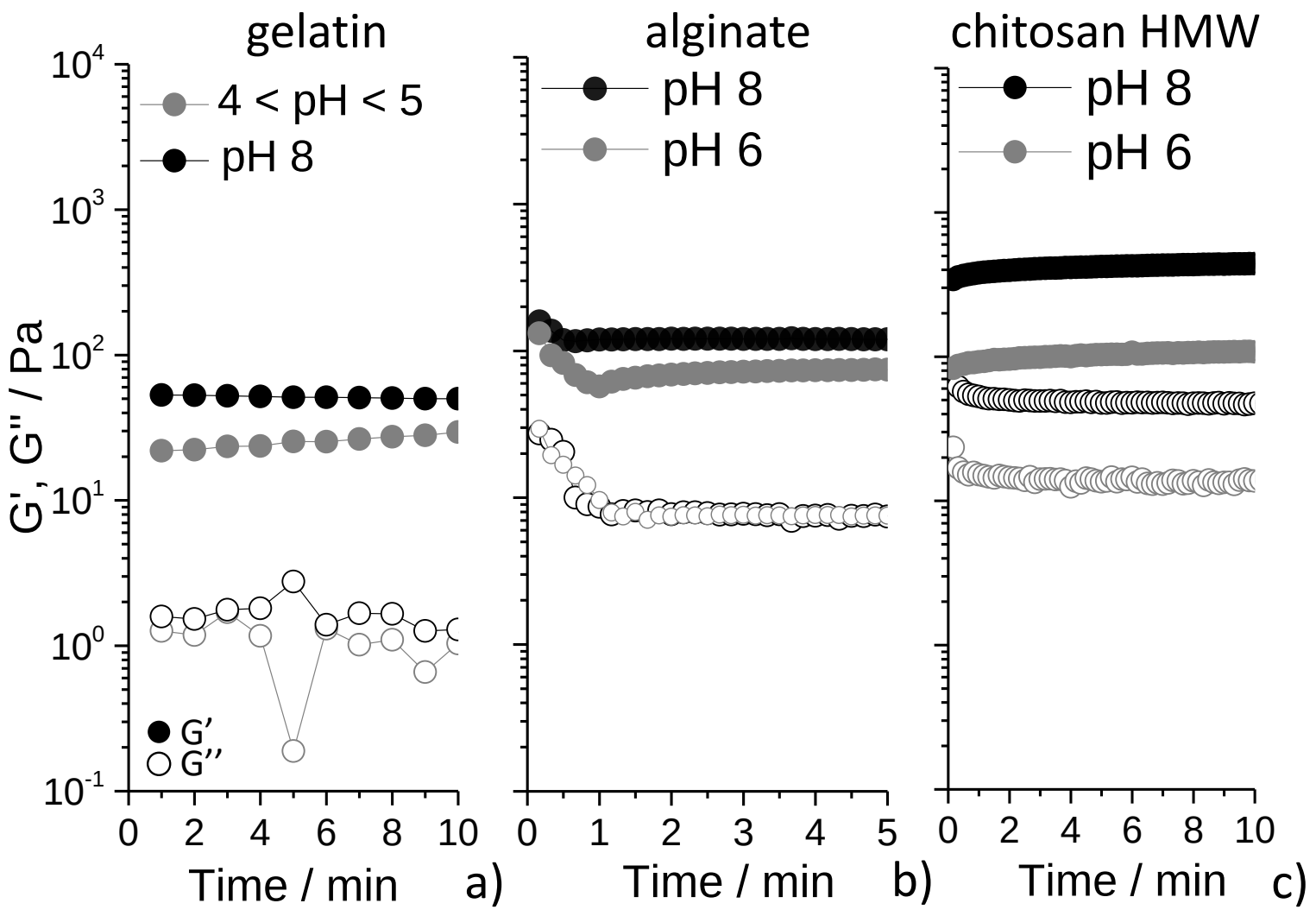

**Figure S 1 – Time-dependence of G' and G'' (LVER, f= 1 Hz, γ= 0.1 %) for a) gelatin (2 wt%), b) alginate (1 wt% [$Ca^{2+}$]= 25 mM) and c) chitosan HMW (1 wt%) control hydrogels at basic and acidic pH. Values of pH are given on top of each panel.**

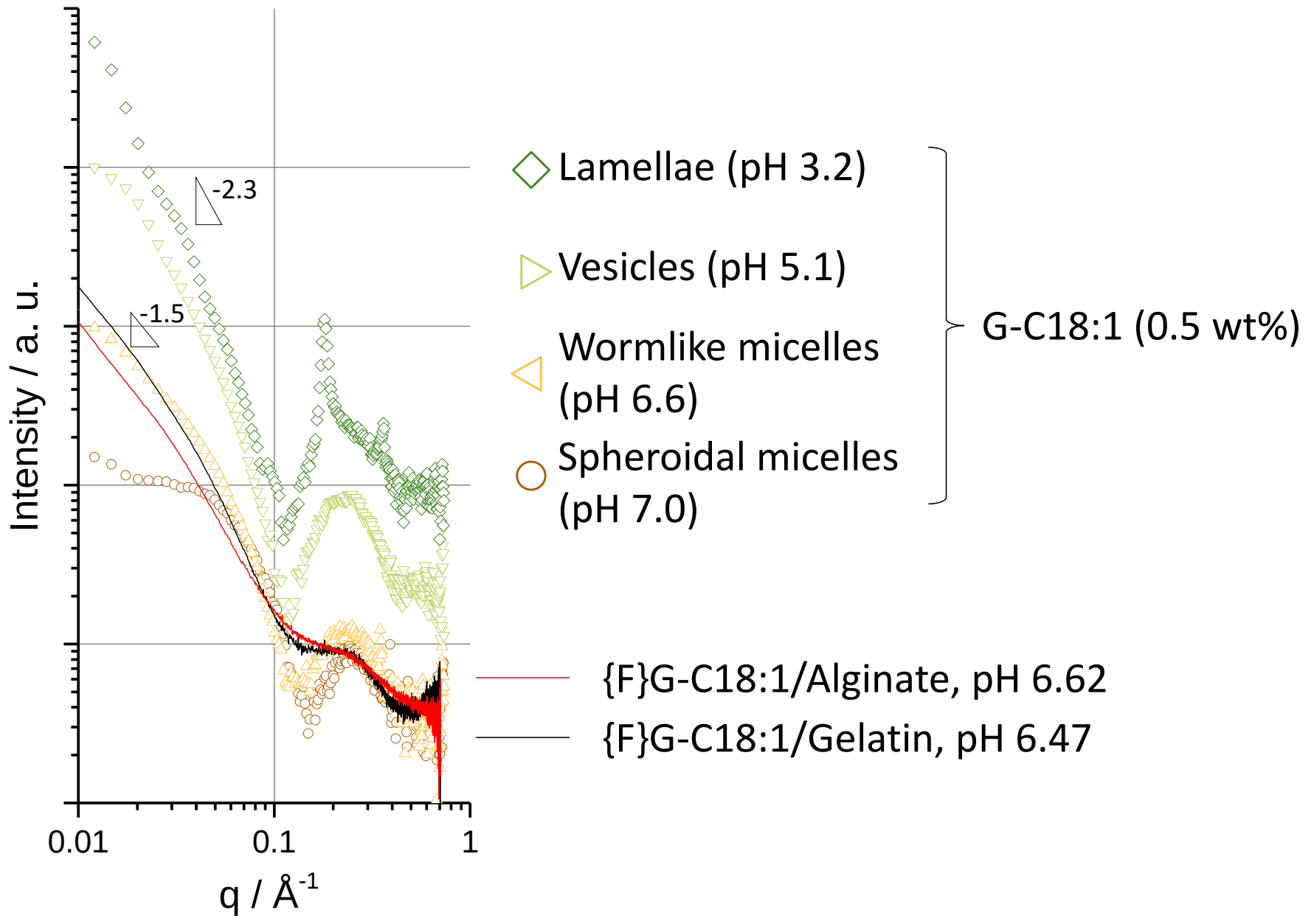

a)

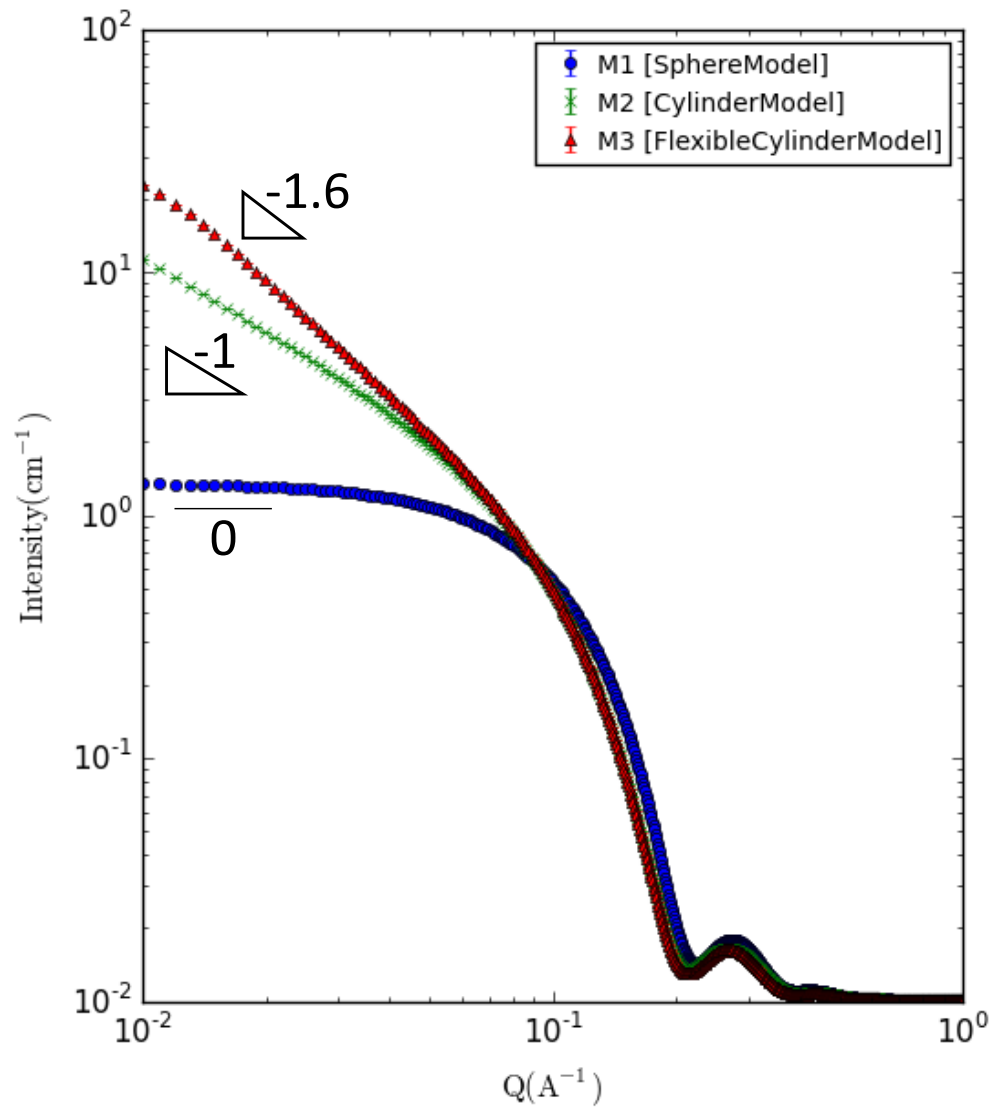

b)

**Figure S 2 – a) Comparison between the SAXS profiles of {F}G-C18:1/Gelatin at pH 6.47 (black line, extracted from Figure 1b in the main text), {F}G-C18:1/Alginate at pH 6.62 (red line, extracted from Figure 2b in the main text) and an aqueous solution (0.5 wt%) of G-C18:1 prepared at various pH values. The data from the latter have been extracted and adapted from Langmuir, *2016*, 32, 6343. For readability purposes, intensity scaling of the SAXS profiles is arbitrary. b) SAXS profiles generated from a sphere, cylinder and flexible cylinder model using the SasView v.3.1.2 software. Highlight is on the typical wavevector (q) dependency of the intensity (slope), being 0 for spheres, -1 for cylinders and -1.6 for flexible cylinders. The parameters of the models are: radius= 20 Å (sphere), 18 Å (cylinder, flexible cylinder), scale= 1 (all), background= 0.01 (all), scattering length density (sld) solvent (value is calculated for $H_2O$)= $9.4 \cdot 10^{-6}$ Å$^{-2}$ (all), sld core (value is typical for a hydrated organic molecule)= $10 \cdot 10^{-6}$ Å$^{-2}$ (all), length= 1000 Å (cylinder, flexible cylinder), Kuhn length= 100 Å (cylinder, flexible cylinder).**

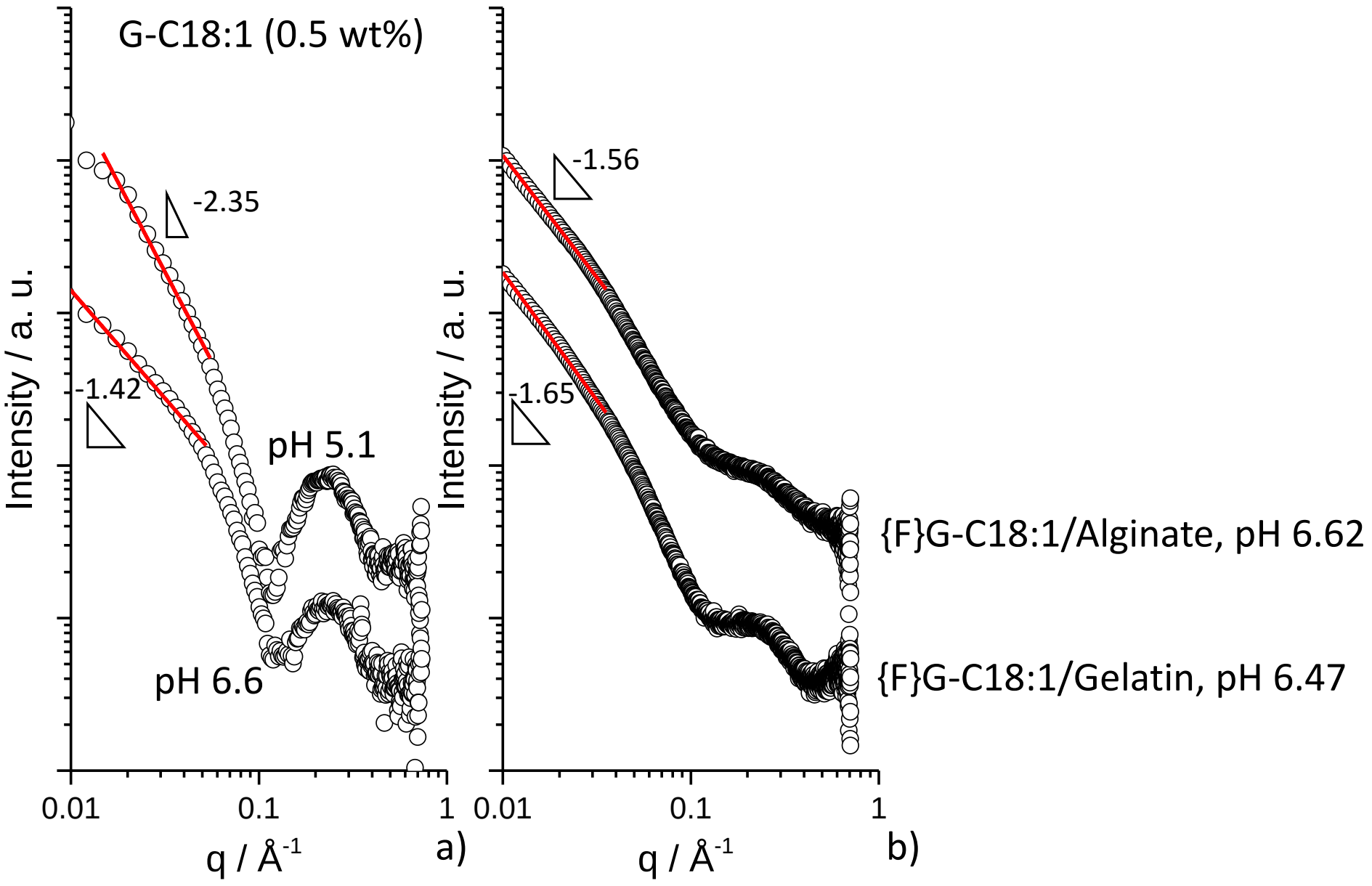

**Figure S 3 – Low-q scaling of the SAXS scattered intensity for a) G-C18:1 and b) {F}G-C18:1/biopolymer samples.**

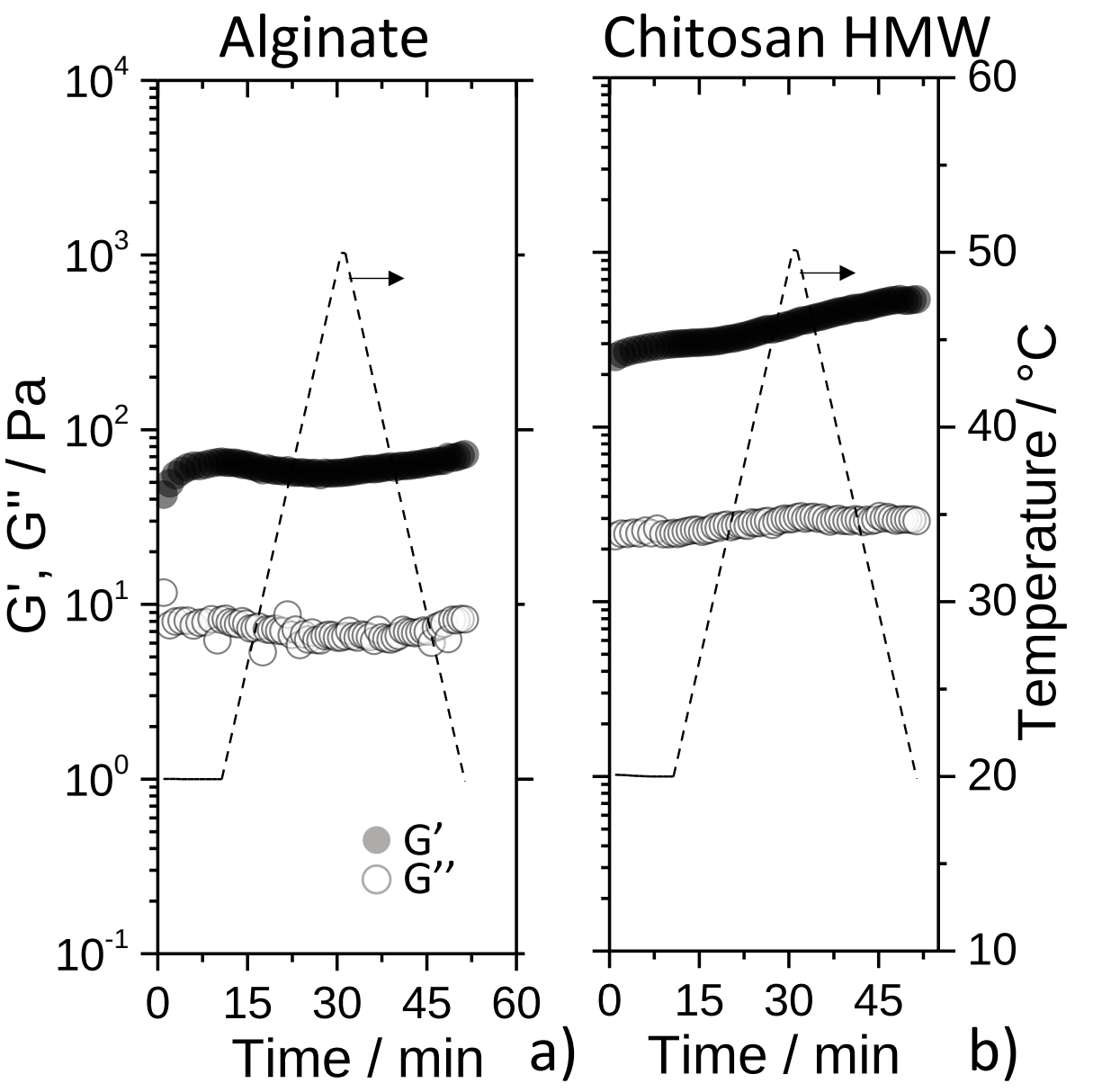

**Figure S 4 – Temperature dependence of G' and G'' (LVER, f= 1 Hz, γ= 0.1%) recorded for a) alginate (1 wt%, pH= 8, [$Ca^{2+}$]= 25 mM) and b) chitosan HMW (1 wt%, pH 8) control hydrogels.**

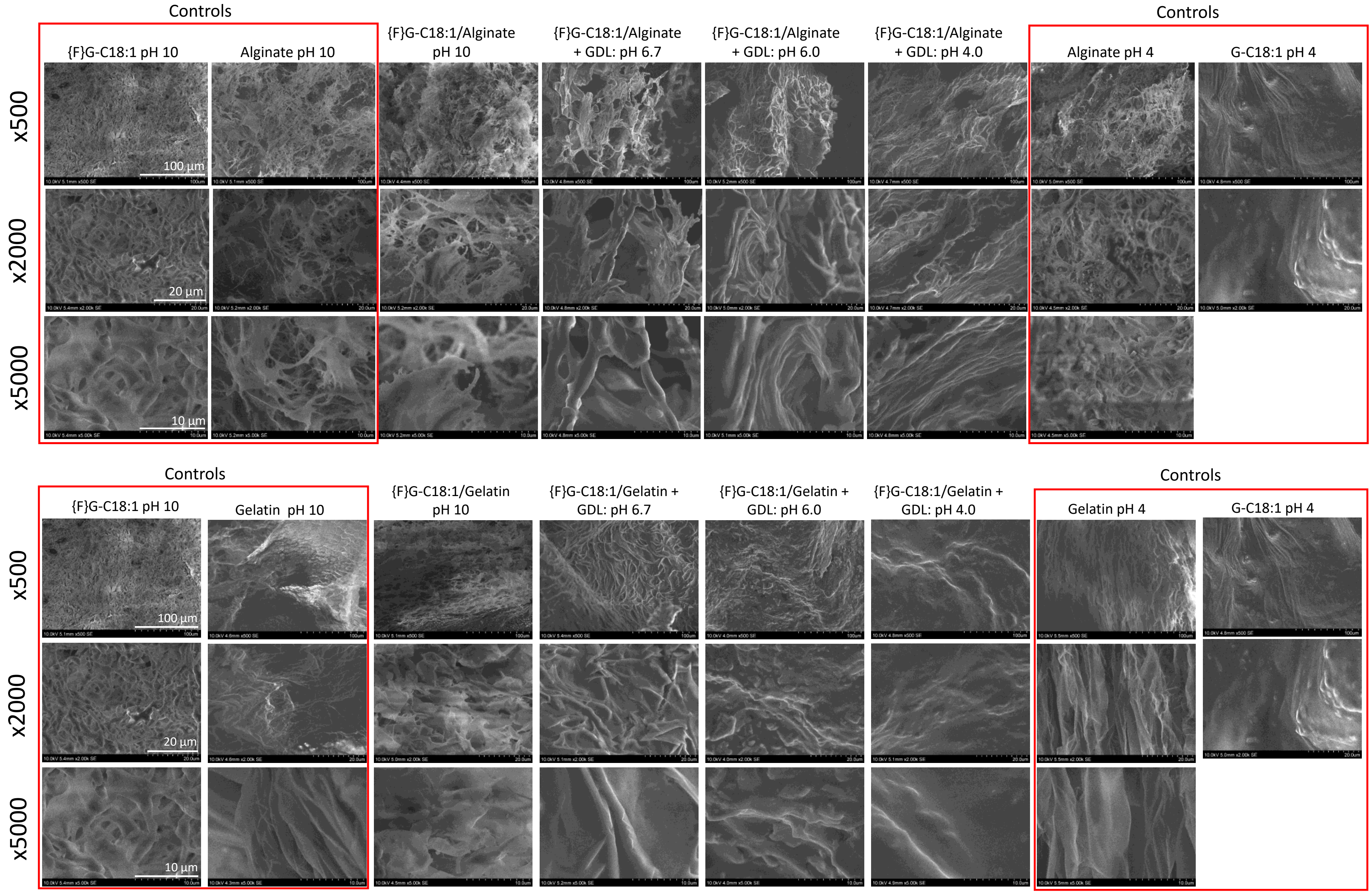

**Figure S 5 – SEM images collected for freeze-dried hydrogels prepared from: {F}G-C18:1, gelatin, alginate, {F}G-C18:1/biopolymer. Sample preparation is given in the materials and methods section of the main manuscript.**

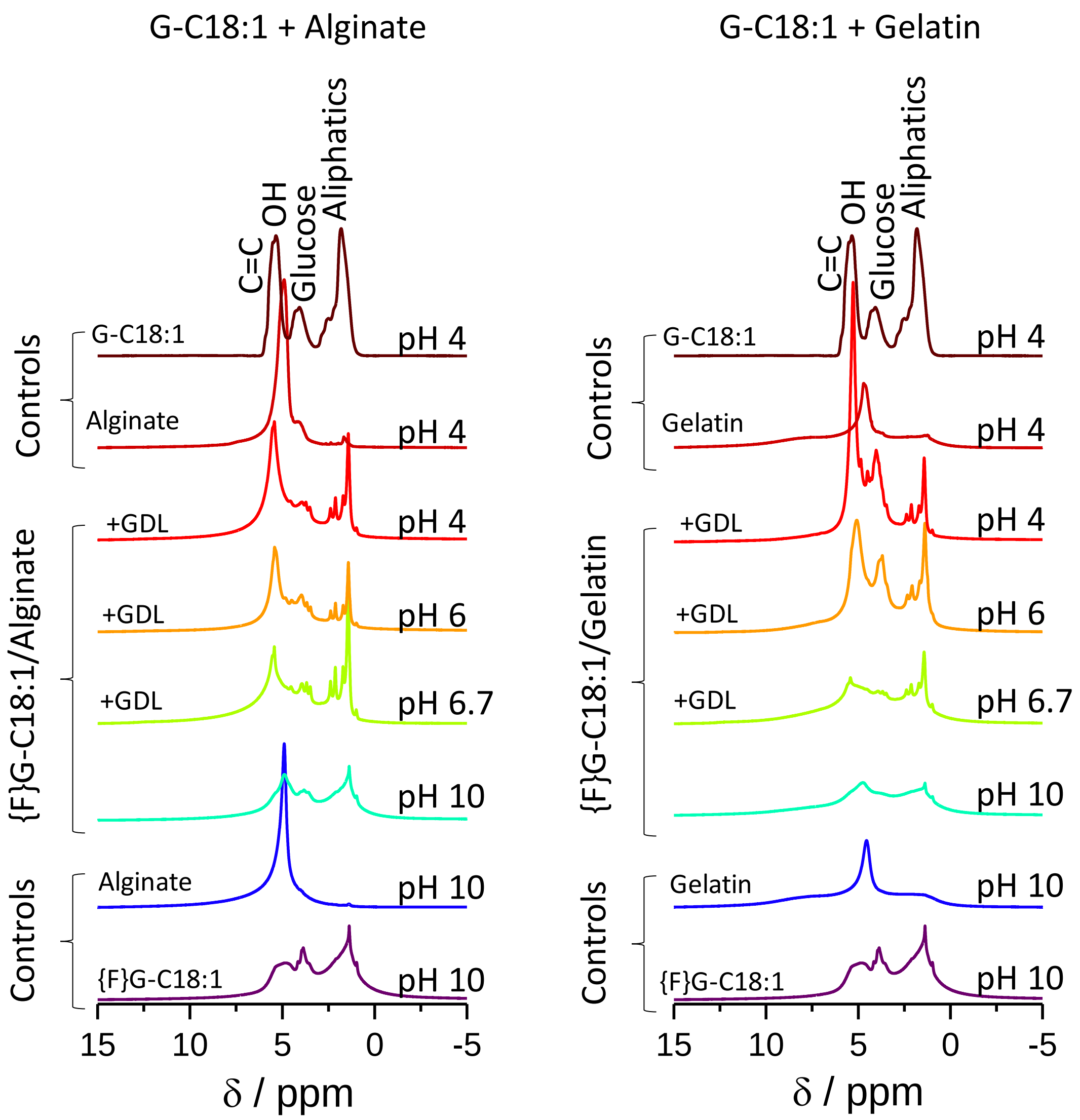

**Figure S 6 – $^{1}H$ and $^{13}C$ solid-state MAS NMR spectra collected for freeze-dried hydrogels prepared from: {F}G-C18:1, gelatin, alginate, {F}G-C18:1/biopolymer. Sample preparation is given in the materials and methods section of the main manuscript. Attribution of the $^{1}H$ spectra can be found in Baccile *et al.*, pH-Driven Self-Assembly of Acidic Microbial Glycolipids. *Langmuir* 2016, 32, 6343. Attribution of $^{13}C$ spectra can be found in Baccile *et al.*, Chameleonic Amphiphile: The Unique Multiple Self-Assembly Properties of a Natural Glycolipid in Excess of Water. *J. Colloids Interface Sci.* 2022, DOI: 10.1016/j.jcis.2022.07.130. The broad peak in the $1^{3}C$ spectra marked by the symbol * is an artifact due to the signal of the probehead.**

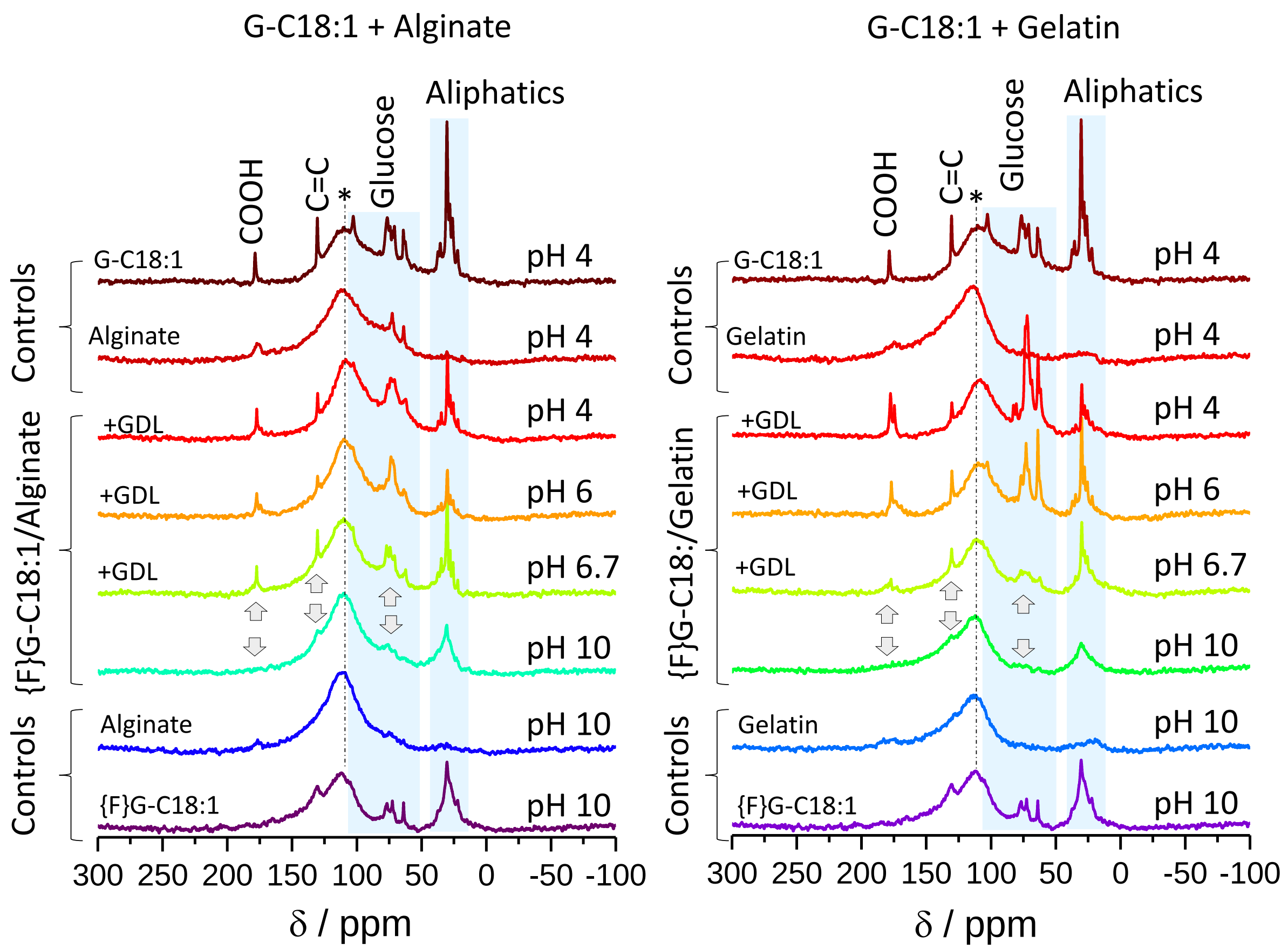

**Figure 6 (continued) – [1]H and [13]C solid-state MAS NMR spectra collected for freeze-dried hydrogels prepared from: {F}G-C18:1, gelatin, alginate, {F}G-C18:1/biopolymer. Sample preparation is given in the materials and methods section of the main manuscript. Attribution of the [1]H spectra can be found in Baccile *et al.*, pH-Driven Self-Assembly of Acidic Microbial Glycolipids. *Langmuir* 2016, 32, 6343. Attribution of [13]C spectra can be found in Baccile *et al.*, Chameleonic Amphiphile: The Unique Multiple Self-Assembly Properties of a Natural Glycolipid in Excess of Water. *J. Colloids Interface Sci.* 2022, DOI: 10.1016/j.jcis.2022.07.130. The broad peak in the [13]C spectra marked by the symbol * is an artifact due to the signal of the probehead.**